\def\kms    {\ifmmode{{\rm \ts km\ts s}^{-1}}\else{\ts km\ts s$^{-1}$}\fi}
\def\msol   {\ifmmode{{\rm M}_{\odot} }\else{M$_{\odot}$}\fi}
\def\lsol   {\ifmmode{L_{\odot}}\else{$L_{\odot}$}\fi}
\def\lfir   {\ifmmode{L_{\rm FIR}}\else{$L_{\rm FIR}$}\fi}
\def\lir   {\ifmmode{L_{\rm IR}}\else{$L_{\rm IR}$}\fi}
\def\zsol   {\ifmmode{{\rm Z}_{\odot}}\else{Z$_{\odot}$}\fi}
\def\etal   {{\rm et\ts al.}}

\def\ts     {\thinspace}
\def\ci     {\ifmmode{[{\rm C}{\rm \small I}]}\else{[C\ts {\scriptsize I}]}\fi}
\def\cii    {\ifmmode{[{\rm C}{\rm \small II}]}\else{[C\ts {\scriptsize II}]}\fi}
\def\nii    {\ifmmode{[{\rm N}{\rm \small II}]}\else{[N\ts {\scriptsize II}]}\fi}
\def\mue {\ifmmode{\mu{\rm m}}\else{$\mu$m}\fi}

\documentclass[iop,numberedappendix]{emulateapj}

\usepackage{natbib}
\usepackage{here}
\usepackage{epsf}
\usepackage{epstopdf}
\usepackage{multirow}
\usepackage{array}

\usepackage{color}

\shorttitle{The SPT-DSFG redshift distribution}
\shortauthors{Strandet, M. et al.}

\begin{document}

\title{THE REDSHIFT DISTRIBUTION OF DUSTY STAR FORMING GALAXIES FROM THE SPT SURVEY}
\def\MPIfR{1}\def\IMPRS{2}

\def\MPIfR{1}
\def\IMPRS{2}
\def\Illinois{3}
\def\ESOGarching{4}
\def\UPenn{5}
\def\Diego{6}
\def\CfA{7}
\def\JPL{8}
\def\KICPChicago{9}
\def\PhysicsUChicago{10}
\def\EFIChicago{11}
\def\AAUChicago{12}
\def\Dal{13}
\def\Colorado{14}
\def\Davis{15}
\def\UFlorida{16}
\def\UCL{17}
\def\Stanford{18}
\def\Arizona{19}
\def\UCLA{20}
\def\IPAC{21}
\def\Oxford{22}

\author{
M.~L.~Strandet$^{\MPIfR,\IMPRS}$,
%
%
A.~Wei\ss$^{\MPIfR}$,
J.~D.~Vieira$^{\Illinois}$,
C.~de~Breuck$^{\ESOGarching}$,
%
%
J.~E.~Aguirre$^{\UPenn}$,
M.~Aravena$^{\Diego}$,
M.~L.~N.~Ashby$^{\CfA}$,
M.~B\'ethermin$^{\ESOGarching}$,
C.~M.~Bradford$^{\JPL}$,
J.~E.~Carlstrom$^{\KICPChicago,\PhysicsUChicago,\EFIChicago,\AAUChicago}$, 
S.~C.~Chapman$^{\Dal}$,
T.~M.~Crawford$^{\KICPChicago,\AAUChicago}$, 
W.~Everett$^{\Colorado}$,
C.~D.~Fassnacht$^{\Davis}$,
R. ~M.~Furstenau$^{\Illinois}$,
A.~H.~Gonzalez$^{\UFlorida}$, 
T.~R.~Greve$^{\UCL}$,	
B.~Gullberg$^{\ESOGarching}$, 
Y.~Hezaveh$^{\Stanford}$,
J.~R.~Kamenetzky$^{\Arizona}$,
K.~Litke$^{\Arizona}$,
J.~Ma$^{\UFlorida}$, 
M.~Malkan$^{\UCLA}$,
D.~P.~Marrone$^{\Arizona}$,   
K.~M.~Menten$^{\MPIfR}$
E.~J.~Murphy$^{\IPAC}$,
A.~Nadolski$^{\Illinois}$,
K.~M.~Rotermund$^{\Dal}$,
J.~S.~Spilker$^{\Arizona}$,
A.~A.~Stark$^{\CfA}$, 
N.~Welikala$^{\Oxford}$
}

\altaffiltext{\MPIfR}{Max-Planck-Institut f\"{u}r Radioastronomie, Auf dem H\"{u}gel 69 D-53121 Bonn, Germany}
\altaffiltext{\IMPRS}{Member of the International Max Planck Research School (IMPRS) for Astronomy and Astrophysics at the Universities of Bonn and Cologne}
\altaffiltext{\ESOGarching}{European Southern Observatory, Karl Schwarzschild Stra\ss e 2, 85748 Garching, Germany}
\altaffiltext{\Illinois}{Department of Astronomy and Department of Physics, University of Illinois, 1002 West Green St., Urbana, IL 61801}
\altaffiltext{\UPenn}{University of Pennsylvania, 209 South 33rd Street, Philadelphia, PA 19104, USA}
\altaffiltext{\Diego}{N\'ucleo de Astronom\'{\i}a, Facultad de Ingenier\'{\i}a, Universidad Diego Portales, Av. Ej\'ercito 441, Santiago, Chile}
\altaffiltext{\CfA}{Harvard-Smithsonian Center for Astrophysics, 60 Garden Street, Cambridge, MA 02138, USA}
\altaffiltext{\JPL}{Jet Propulsion Laboratory, 4800 Oak Grove Drive, Pasadena, CA 91109, USA}
\altaffiltext{\KICPChicago}{Kavli Institute for Cosmological Physics, University of Chicago, 5640 South Ellis Avenue, Chicago, IL 60637, USA}
\altaffiltext{\PhysicsUChicago}{Department of Physics, University of Chicago, 5640 South Ellis Avenue, Chicago, IL 60637, USA}
\altaffiltext{\EFIChicago}{Enrico Fermi Institute, University of Chicago, 5640 South Ellis Avenue, Chicago, IL 60637, USA}
\altaffiltext{\AAUChicago}{Department of Astronomy and Astrophysics, University of Chicago, 5640 South Ellis Avenue, Chicago, IL 60637, USA}
\altaffiltext{\Dal}{Dalhousie University, Halifax, Nova Scotia, Canada}
\altaffiltext{\Colorado}{Department of Astrophysical and Planetary Sciences and Department of Physics, University of Colorado, Boulder, CO 80309}
\altaffiltext{\Davis}{Department of Physics,  University of California, One Shields Avenue, Davis, CA 95616, USA}
\altaffiltext{\UFlorida}{Department of Astronomy, University of Florida, Gainesville, FL 32611, USA}
\altaffiltext{\UCL}{Department of Physics and Astronomy, University College London, Gower Street, London WC1E 6BT, UK}
\altaffiltext{\Stanford}{Kavli Institute for Particle Astrophysics and Cosmology, Stanford University, Stanford, CA 94305, USA}
\altaffiltext{\Arizona}{Steward Observatory, University of Arizona, 933 North Cherry Avenue, Tucson, AZ 85721, USA}
\altaffiltext{\UCLA}{Department of Physics and Astronomy, University of California, Los Angeles, CA 90095-1547, USA}
\altaffiltext{\IPAC}{Infrared Processing and Analysis Center, California Institute of Technology, MC 220-6, Pasadena, CA 91125, USA}
\altaffiltext{\Oxford}{Department of Physics, Oxford University, Denys Wilkinson Building, Keble Road, Oxford, OX1 3RH, UK}

\begin{abstract}											 %
We use the Atacama Large Millimeter/submillimeter Array (ALMA) in Cycle 1 to determine spectroscopic redshifts of high-redshift dusty star-forming galaxies (DSFGs) selected by their 1.4\,mm continuum emission in the South Pole Telescope (SPT) survey.  
We present ALMA 3\,mm spectral scans between 84-114\,GHz for 15 galaxies and targeted ALMA 1\,mm observations for an additional eight sources. 
Our observations yield 30 new line detections from CO, \ci , \nii , H$_2$O and NH$_3$.  
We further present APEX \cii\ and CO mid-$J$ observations for seven sources for which only a single line was detected in spectral-scan data from ALMA Cycle 0 or Cycle 1.
We combine the new observations with previously published and new mm/submm line and photometric data of the SPT-selected DSFGs to study their redshift distribution.  
The combined data yield 39 spectroscopic redshifts from molecular lines, a success rate of $>$$85\%$.
Our sample represents the largest data set of its kind today and has the highest spectroscopic completeness among all redshift surveys of high-$z$ DSFGs. 
The median of the redshift distribution is $z$$=$$3.9\pm 0.4$, and the highest-redshift source in our sample is at $z$$=$$5.8$. 
We discuss how the selection of our sources affects the redshift distribution, focusing on source brightness, selection wavelength, and strong gravitational lensing.
We correct for the effect of gravitational lensing and find the redshift distribution for 1.4\,mm-selected sources with a median redshift of $z$$=$$3.1\pm 0.3$. 
Comparing to redshift distributions selected at shorter wavelengths from the literature, we show that selection wavelength affects the shape  of the redshift distribution.
\end{abstract}


\section{Introduction} 										 %

In the last two decades, millimeter (mm) and submillimeter (submm) surveys have transformed our understanding of galaxy formation and evolution by revealing that luminous, dusty galaxies were a thousand times more abundant in the early Universe than they are at the present day \citep[e.g., see review by][]{casey14}. 
The first spectroscopic redshift distributions of submm-selected galaxies indicated that the population of dusty star forming galaxies (DSFGs) peaked at redshift $z $$\sim$$2.3$ \citep[e.g.,][]{chapman05}, coeval with the peak of black hole accretion and cosmic star formation \citep[e.g.,][]{hopkins06}. 
These studies suggested that a significant fraction of star formation activity in the universe at $z$$=$$2$$-$$3$ is taking place in DSFGs brighter than ${\rm S_{850\mu m}}$$\approx$$1$\,mJy, and could be hidden from the view of optical/UV observations owing to the large dust obscuration \citep[e.g.,][]{wardlow11}. 
Theoretical models suggest that the contribution of DSFGs to the total star formation rate density at $z=2--4$  is of order 10\% \citep[for sources with $S_{870\mu m}$$>$1\,mJy;][]{gonzalez11}.
While the history of star formation has now been measured out to $z\sim8$ through rest-frame UV surveys \citep[see review by][]{madau14}, progress in measuring highly obscured star formation as a function of look-back time has been much slower, mainly because of the difficulties in obtaining robust redshifts for DSFGs.

Dust emission at high redshift ($z$$>$$1$) exhibits a steep rise on the Rayleigh-Jeans side of the greybody spectrum that counteracts the dimming from luminosity distance \citep[][]{blain93}.
This very negative K-correction is sufficient to produce a nearly redshift-independent selection of DSFGs at mm/submm wavelengths.

However, the poor spatial resolution ($\sim$$20''$) of the single-dish submm telescopes used to perform extra-galactic surveys has prevented immediate counterpart identification. 
This difficulty was further compounded by the dust-obscured nature of DSFGs, which makes counterpart identification at optical wavelengths difficult or impossible.
Even with high spatial resolution data taken at radio \citep[e.g.,][]{ivison02} and/or mid-infrared \citep[e.g.,][]{ashby06, pope06} wavelengths, the slope of the spectral energy distributions (SEDs) of galaxies in the radio or mid-infrared (MIR) is such that the K-correction is positive, and galaxies become more difficult to detect at high redshifts.  Thus,  50\% of DSFGs typically lack robust counterparts at other wavelengths \citep[e.g.,][]{biggs11}, although the exact fraction depends on the depth of the radio/MIR observations. This mismatch in sensitivity at different wavelengths has potentially left the highest-redshift sources ($z$$>$3) unidentified, which would bias the observed redshift distribution of DSFGs low.

Millimeter interferometry provides a more reliable and complete method to obtain secure multi-wavelength identifications of DSFGs discovered in  single-dish surveys. 
\citet{Dannerbauer02} first published counterpart identifications based on high spatial resolution data for three 1.2\,mm-selected DSFGs observed with the IRAM Plateau de Bure Interferometer (PdBI), and  \citet{younger07}  used the Submillimeter Array to identify counterparts to seven 1.1\,mm-detected sources. In spite of the accurate and reliable positions, neither study  successfully obtained redshifts for the DSFGs, although one of the sources was eventually determined to be at a record breaking (for DSFGs) $z$$=$5.3 from rest frame UV spectroscopy \citep{riechers10, capak11}.
\citet{smolcic12} used PdBI to follow up a sample of 1.1\,mm selected DSFGs, leading to optical spectroscopic redshifts for roughly half the sample and photometric redshift estimates for the remaining sources, and these data suggested that the previous spectroscopically determined redshift distributions of DSFGs \citep[e.g.,][]{chapman05} were biased low. 
Other follow-up efforts have led to different conclusions. For example, \citet{simpson14} and more recently \citet{dacunha15} use the 17-band optical to mid-IR photometry of the Extended Chandra Deep Field South (ECDFS) to study the photometric redshift distribution of DSFGs with counterpart identification based on high resolution Atacama Large Millimeter/submillimeter Array (ALMA) 870\mue\ observations \citep{hodge13}. \citet{simpson14} derive a median redshift of $\bar{z}$=2.5, albeit with a significant 
tail of DSFGs at $z$$>$4.  This result is consistent with the early findings of \citet{chapman05} under the assumption that \citet{chapman05} did not detect the high-redshift tail since that study only targeted radio-confirmed DSFGs.

In the past few years, new instruments with larger bandwidths have enabled a more direct and unbiased way to derive redshifts of DSFGs via observations of molecular emission lines at mm wavelengths.
The molecular line emission, typically from CO, or \cii,  can be related unambiguously to the mm/submm dust continuum, circumventing the need for high-resolution imaging, counterpart identification, and optical spectroscopy  \citep{weiss09,harris12,lupu12,walter12,chapman15}. 
The first redshift distribution based on molecular emission lines detected via blind spectral scans in the 3\,mm window using ALMA was published by \citet{weiss13} for 26 strongly lensed sources selected from the 2500 degree$^2$ South Pole Telescope (SPT) survey.
Performing a redshift search for such a big sample in the early stage of ALMA operations was only possible due to the strongly lens-magnified nature of the sources which makes them extraordinarily bright. The redshift distribution of the SPT sample has a much higher mean ($\bar{z}$=3.5) than observed for any other sample of DSFGs and has stimulated an on-going discussion on the redshift distribution of DSFGs in the literature \citep[e.g.,][]{koprowski14,miettinen15,bethermin15}.

Progress has also been made towards a theoretical understanding of the differences seen in observed redshift distributions. Recently \citet{bethermin15} modeled the expected DSFG redshift distribution based on a phenomenological model of galaxy evolution. They conclude the difference can be understood in terms of survey selection wavelength and, to a minor degree, the survey depth. 
In addition, they investigate the effect of gravitational lensing on the redshift distribution.
At wavelengths shorter than 1.1\,mm the lensed redshift distribution always tends to show a higher median redshift than the unlensed distribution. At longer selection wavelengths, as investigated here, the effect of gravitational lensing on the redshift distribution vanishes unless only extremely luminous sources are selected (e.g., $S_{1.4mm}$$>$25\,mJy)

In this paper we extend the ALMA redshift survey of 26 SPT-selected sources from \citet{weiss13} with an additional 15 sources. 
We use this extended sample to construct an updated redshift distribution of SPT-selected DSFGs. We further present data from ALMA and the Atacama Pathfinder Experiment (APEX) used to confirm redshifts 
for a sample of sources with ambiguous redshifts from \citet{weiss13} and the new survey.
In Section \ref{sect:observations}, we present the ALMA observations along with \cii\ and CO observations carried out with APEX. In Section \ref{sect:results}, we show the spectra derived from these observations and present redshifts determined from those spectra. In Section \ref{sect:discussion}, we present the redshift distribution of DSFGs selected from the SPT survey and discuss how the sample is affected by gravitational lensing and selection wavelength.

We adopt a flat $\Lambda$CDM cosmology, with $\Omega_{\Lambda} = 0.696$ and $H_0 = 68.1$ km s$^{-1}$\,Mpc$^{-1}$ \citep{planck14}.

\section{Observations} \label{sect:observations}					 %

\begin{deluxetable}{p{1.7cm} p{3.0cm} p{1.3cm} p{1.4cm}}
	\tabletypesize{\footnotesize}
	\tablecaption{Summary of spectroscopic observations presented in this work, with ALMA 3\,mm continuum positions\label{Tab:ALMApos}}
	\tablewidth{0pt}
	\tablehead{
	\colhead{Short name} & \colhead{Source} & \colhead{RA} & \colhead{Dec}}
	\startdata
\\
\multicolumn{4}{c}{\textbf{ALMA 1\,mm band 6 redshift confirmation (Figure \ref{Fig:spectra_cy1})}}\\[2mm]  
\hline
\rule{0pt}{3ex}\multirow{1}{*}{SPT0125-50\tablenotemark{a}}	& SPT--S J012506-4723.7	& 01:25:07.08	& -50:38:20.9	\\
\rule{0pt}{2.5ex}\multirow{1}{*}{SPT0300-46\tablenotemark{a}}	& SPT--S J030003-4621.3	& 03:00:04.37	& -46:21:24.3	\\
\rule{0pt}{2.5ex}\multirow{1}{*}{SPT0319-47\tablenotemark{a}}	& SPT--S J031931-4724.6	& 03:19:31.88	& -47:24:33.7	\\
\rule{0pt}{2.5ex}\multirow{1}{*}{SPT0441-46\tablenotemark{a}}	& SPT--S J044143-4605.3	& 04:41:44.08	& -46:05:25.5	\\
\rule{0pt}{2.5ex}\multirow{1}{*}{SPT0459-58\tablenotemark{a}}	& SPT--S J045859-5805.1	& 04:58:59.80	& -58:05:14.0	\\
\rule{0pt}{2.5ex}\multirow{1}{*}{SPT0512-59\tablenotemark{a}}	& SPT--S J051258-5935.6	& 05:12:57.98	& -59:35:41.9	\\
\rule{0pt}{2.5ex}\multirow{1}{*}{SPT0550-53\tablenotemark{a}}	& SPT--S J055001-5356.5	& 05:50:00.56	& -53:56:41.7	\\
\rule{0pt}{2.5ex}\multirow{1}{*}{SPT2132-58\tablenotemark{a}}	& SPT--S J213242-5802.9	& 21:32:43.23	& -58:02:46.2	\\
\\  
\hline 
\\
\multicolumn{4}{c}{\textbf{ALMA 3\,mm band 3 redshift search (Figure \ref{Fig:spectra})}}\\[2mm] 
\hline
\rule{0pt}{3ex}SPT0002-52							& SPT--S J000223-5232.1	& 00:02:23.24	& -52:31:52.5	\\
\rule{0pt}{2.5ex}SPT2307-50							& SPT--S J230726-5003.8	& 23:07:24.71	& -50:03:35.6	\\
\rule{0pt}{2.5ex}SPT2311-54							& SPT--S J231125-5450.5	& 23:11:23.94	& -54:50:30.0	\\
\rule{0pt}{2.5ex}SPT2319-55							& SPT--S J231922-5557.9	& 23:19:21.67	& -55:57:57.8	\\
\rule{0pt}{2.5ex}SPT2335-53							& SPT--S J233513-5324.0	& 23:35:13.15	& -53:24:29.9	\\
\rule{0pt}{2.5ex}\multirow{1}{*}{SPT2340-59\tablenotemark{b}} & SPT--S J234009-5943.1	& 23:40:09.36	& -59:43:32.8	\\
												&						& 23:40:08.95	& -59:43:32.0	\\
\rule{0pt}{2.5ex}SPT2349-50							& SPT--S J234942-5053.5	& 23:49:42.16	& -50:53:30.7	\\
\rule{0pt}{2.5ex}\multirow{1}{*}{SPT2349-56\tablenotemark{b}} & SPT--S J234944-5638.3	& 23:49:42.68	& -56:38:19.4	\\
 												&						& 23:49:42.79	& -56:38:23.9 	\\
 												&						& 23:49:42.84	& -56:38:25.0 	\\
\rule{0pt}{2.5ex}SPT2351-57							& SPT--S J235149-5722.2	& 23:51:50.79	& -57:22:18.3	\\
\rule{0pt}{2.5ex}SPT2353-50							& SPT--S J235339-5010.1	& 23:53:39.22	& -50:10:08.2	\\
\rule{0pt}{2.5ex}SPT2354-58							& SPT--S J235434-5815.1	& 23:54:34.27	& -58:15:08.4	\\
\rule{0pt}{2.5ex}SPT2357-51							& SPT--S J235718-5153.6	& 23:57:16.84	& -51:53:52.9	\\
 \\ \hline
 \\
\multicolumn{4}{c}{\textbf{APEX/FLASH redshift confirmation (Figure \ref{Fig:CII-lines} and \ref{Fig:line-overlay})}}\\[2mm]  
\hline
\rule{0pt}{3ex}\multirow{1}{*}{SPT0319-47\tablenotemark{a}}	& see above							&			&			\\
\rule{0pt}{2.5ex}\multirow{1}{*}{SPT0551-50\tablenotemark{a}} & SPT--S J055138-5058.0	& 05:51:39.42	& -50:58:02.1	\\
\rule{0pt}{2.5ex}SPT2335-53							& see above				&			&			\\
\rule{0pt}{2.5ex}SPT2349-56							& see above				&			&			\\
\rule{0pt}{2.5ex}SPT2353-50							& see above				&			& 			\\
\\ \hline
\\
\multicolumn{4}{c}{\textbf{APEX/SEPIA redshift confirmation (Figure \ref{Fig:line-overlay})}}\\[2mm]  
\hline
\rule{0pt}{3ex}SPT0002-52							& see above				&			&			\\
\rule{0pt}{3ex}SPT2349-50							& see above				&			&			\\
\\ \hline
\\
\multicolumn{4}{c}{\textbf{APEX/Z-Spec redshift search (Figure \ref{Fig:SPT0551Zspec})}}\\[2mm]  
\hline
\rule{0pt}{3ex}\multirow{1}{*}{SPT0551-48\tablenotemark{c}}	& SPT--S J055156-4825.1	& 05:51:54.65	& -48:25:01.8	\\
	\enddata
	 \tablenotetext{a}{These sources and their positions are from \citet{weiss13}.}
	 \tablenotetext{b}{These sources split into multiple counterparts at 3\,mm; we here give the 870\,\micron\ positions of all counterparts.}
	 \tablenotetext{c}{Position from APEX/LABOCA; No ALMA data.}
\end{deluxetable}

Observations presented in this work include ALMA Cycle 1 observations in the 3~mm and 1~mm bands, 
as well as observations from APEX using the First Light APEX Submillimeter Heterodyne (FLASH) receiver, 
the Swedish-ESO PI receiver for APEX (SEPIA), and the Z-spec camera.

In ALMA Cycle 0, \citet{weiss13} set out to determine redshifts for 26 SPT-selected DSFGs using CO lines in the ALMA 3\,mm band. Unambiguous redshifts (from multiple CO lines) were determined for twelve sources, while 11 sources showed only a single line. 
Using the same strategy in ALMA Cycle 1, we searched for CO in the 3 mm band in 15 new sources; these observations are presented in Section \ref{Sec:ALMA3mm}. 

For sources with only one detected line in either 3~mm redshift search, 
we use well-sampled photometry to determine a photometric redshift and thereby the most probable line 
identification and redshift. We use this information to perform targeted redshift confirmation observations,
either in different ALMA bands or using heterodyne receivers on APEX.
For eight sources with single line detections in the ALMA Cycle 0 redshift search, we obtained ALMA Cycle 1 data in band 6 (1 mm) in an attempt to detect a second CO line or a \nii\ line; we present these observations in Section \ref{Sec:ALMA1mm}. 
For five sources (including one source observed in ALMA band 6 and some single-line detections in the Cycle 1 redshift search), we followed up the most probable redshift options with APEX/FLASH (Section \ref{sect:cii-obs}) and APEX/SEPIA (Section \ref{Sec:SEPIA}). An additional one source in this sample was observed in 2012 with APEX/Z-Spec (Section \ref{Sect:Zspec-obs}).
An overview of these observations is found in Table \ref{Tab:ALMApos}.

\subsection{ALMA 1\,mm follow up observations} \label{Sec:ALMA1mm}

In the ALMA 3\,mm spectral scans presented in \citet{weiss13}, ten sources showed a single CO line detection (plus one source, SPT0319-47, which showed a line feature not significant enough for detection).
In these cases photometric measurements were used to validate possible line assignments and to find the most likely redshift option (this approach is described in more detail in Section \ref{Sect:one-line}).
Using this method the redshifts of three sources were quickly secured by APEX/FLASH follow-up observations in \cii\ \citep{gullberg15}.
For the eight remaining sources, we were awarded observing time with ALMA in the Cycle 1 early science compact array configuration, to search for a second CO line (CO(6--5) -- CO(12--11)) or a \nii\ line in ALMA band 6 (211\,GHz - 275\,GHz) (project ID 2012.1.00994.S). The eight sources observed are listed in Table \ref{Tab:ALMApos}.

The sources were grouped into five science blocks based on their sky position and tuning frequencies of possible redshifted molecular emission lines (mainly CO).
The sidebands were placed so that these five science blocks would yield at least one line for each source.  One source (SPT0441-46) is observed in two tunings since it had two likely redshift options.

The observations were carried out from 2013 December to 2014 December. 
The flux density calibration was based on observations of the Solar System objects Uranus, Neptune and Ganymede and the quasars J0334-401 and J0519-454. The bandpass and phase calibration were determined using nearby quasars.
The number of antennas used during the observations ranged from 25 to 40, with baselines less than 500\,m resulting in a synthesized beam size of 1.5$\times$0.8\,arcsec ($''$). 
In band 6 the primary beam is $29\arcsec$$-$$23\arcsec$.
The observing time for each science block ranged from 8 to 20 minutes on-source, excluding overheads. 
Typical single-sideband (SSB) system temperatures for the observations were $T_{\rm sys}$=80--100\,K. 
The data were processed using the Common Astronomy Software Application package \citep[CASA][]{mcmullin07, petry12}.  We used natural weighting and constructed the spectra with a channel width of 19.5\,MHz ($18$$-$$22$\kms\ for the highest and lowest observing frequency).
The typical noise per channel is 0.9--1.9\,mJy beam$^{-1}$. Continuum images were cleaned and generated from the full bandwidth have typical noise levels of 50\,$\mu$Jy beam$^{-1}$.

\subsection{ALMA 3\,mm scans} \label{Sec:ALMA3mm}
Also in ALMA Cycle 1, we extended the Cycle 0 redshift search from \citet{weiss13} to 15 additional 
SPT-selected DSFGs (project ID 2012.1.00844.S). As in the Cycle 0 observations, we searched for CO lines 
in the 3\,mm atmospheric transmission window (ALMA band 3). 

\begin{figure}[t]
	\centering
	\includegraphics[viewport= 0 0 430 360, clip=true,width=7.5cm,angle=0]{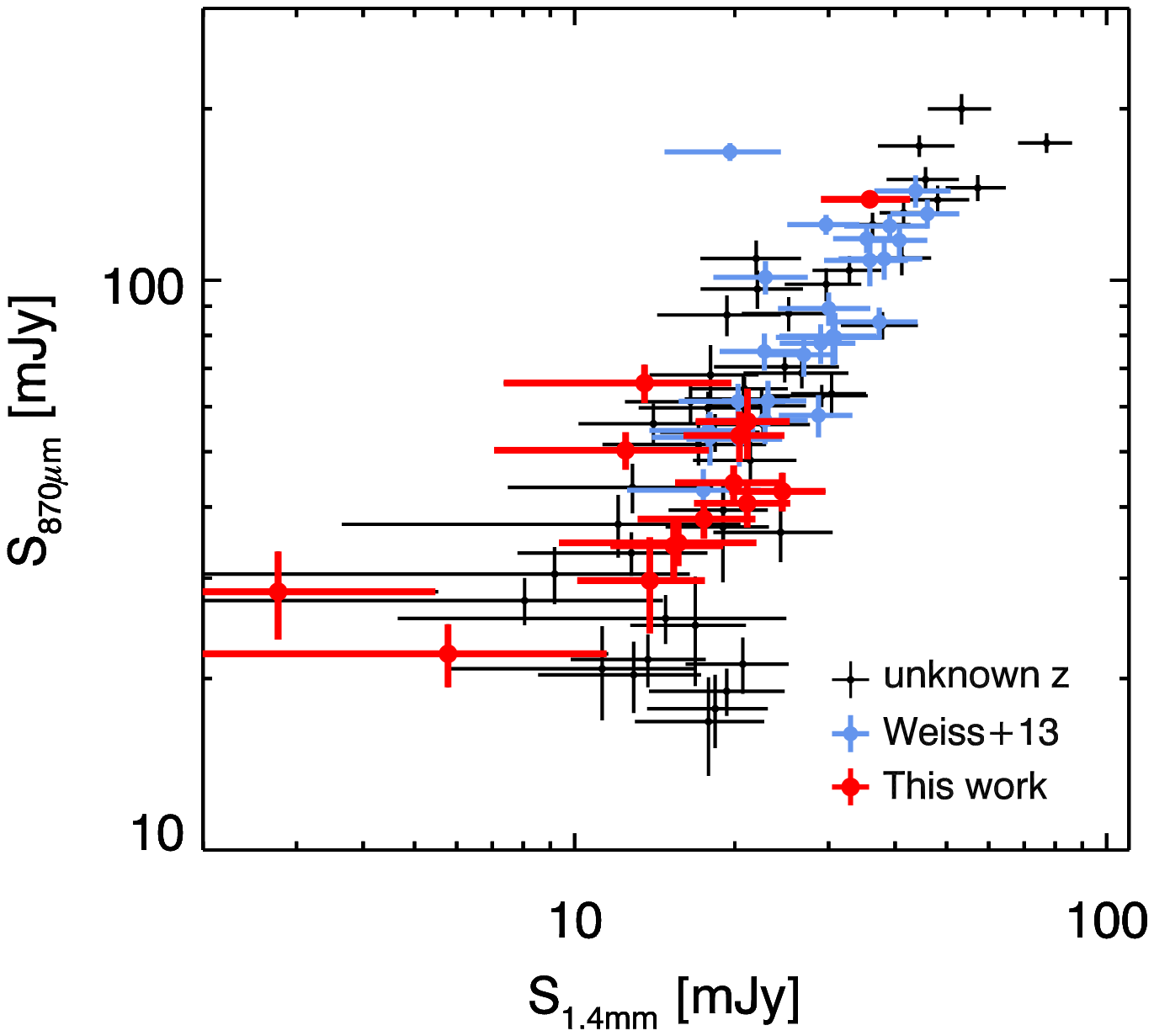}
	\includegraphics[viewport= 0 0 430 360, clip=true,width=7.5cm,angle=0]{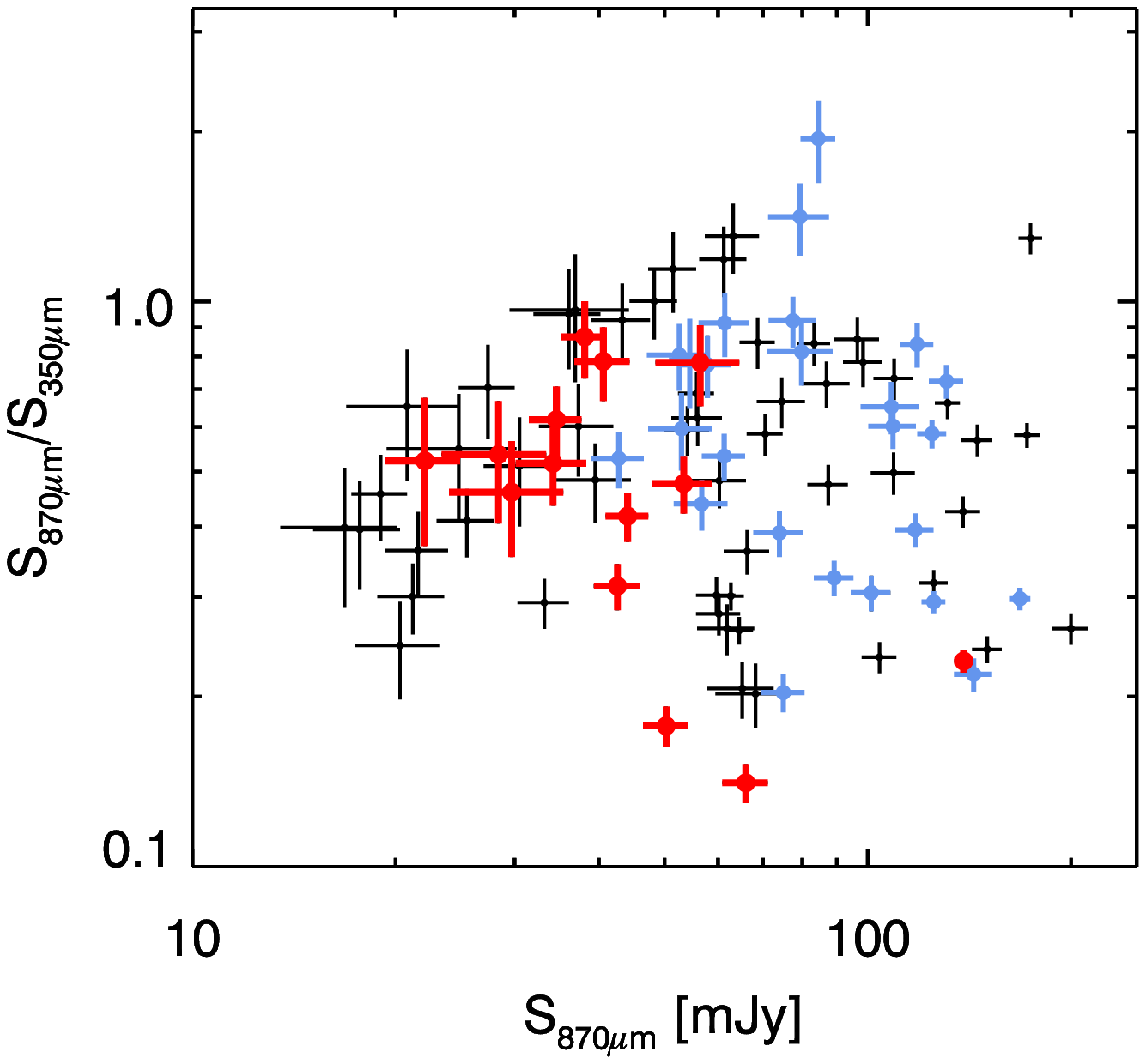}
	\caption{Flux density and color plots for all sources in the SPT-DSFG sample (\emph{black}), with the 28 sources from \citet{weiss13} (\emph{blue}) and the new DSFGs from this work (\emph{red}). Together these two samples constitute a representative subset of the overall SPT-DSFG sample.
{\it Top:} APEX/LABOCA 870\,$\mu$m flux density as a function of SPT 1.4\,mm flux density. {\it Bottom:} The ratio of {\it Herschel}/SPIRE 350\,\micron\ flux density to APEX/LABOCA 870\,\micron\ flux density as a function of APEX/LABOCA 870\,$\mu$m flux density. This color indicates the redness and thereby redshift of the sample.
}
	\label{Fig:sample}
\end{figure}

\begin{figure}[t]
	\centering
	\includegraphics[viewport= 15 30 320 495, clip=true,width=6.0cm,angle=0]{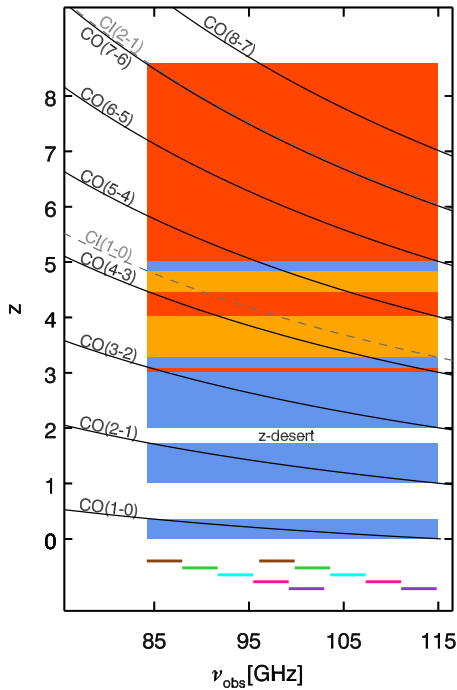}
	\caption{The redshift as a function of the ALMA 3\,mm spectral coverage of the CO and \ci\ emission lines. The \emph{red} shaded area shows redshift ranges where we have two CO lines, the \emph{orange} shaded area shows the redshift ranges where the second line is the weaker \ci\ line and the \emph{blue} shaded area shows the redshift ranges where we will see a single line. At the bottom of the plot the placement of the five tunings are shown.}
	\label{Fig:CO-coverage}
\end{figure}

As in \citet{weiss13}, the sources targeted in the Cycle 1 observations are a subset of a population of rare and extremely bright galaxies in the SPT survey \citep{mocanu13}.  
The sources are selected from 2.0\,mm and 1.4\,mm maps, and further vetoes \citep[described in][]{vieira10, weiss13} remove synchrotron-dominated blazars and low redshift sources ($z$$<$0.1).  The remaining sources are SPT-discovered DSFGs (SPT-DSFGs), which are found to be at high redshift and predominantly strongly lensed \citep{vieira13, hezaveh13, spilker15}. 

The sources in the sample studied here are in the SPT Deep Field that has full coverage from both {\it Herschel}/SPIRE and {\it Spitzer}/IRAC. The sources form a complete, flux-density-limited sample with raw $S_{1.4mm}$$>$16\,mJy within a 10\,$\times$\,10~square-deg field (12 sources). This selection and limitation is a consequence of ALMA Cycle 1 target restrictions requiring that all sources be within 10 deg of each other to share a phase calibrator. In addition we included three fainter sources from the same field (raw $S_{1.4mm}$$\sim$15\,mJy) to reach the maximum number of 15 science targets allowed in this observing setup in Cycle 1.

\citet{weiss13} studied a subset of the SPT-DSFG population selected using a higher raw flux density cut of $S_{1.4mm}$$>$25\,mJy, and thereby picking out the brightest sources of the SPT-DSFG population from a larger area of the sky (1300\,deg$^2$ compared to 100\,deg$^2$ in this work).  
In Figure\,\ref{Fig:sample} we show photometric flux densities and 870\,$\mu$m/350\,$\mu$m flux density ratios for the entire
sample of SPT-DSFGs (\emph{black}), including the sample from this work
(\emph{red}) and the sample targeted in \citet{weiss13} (\emph{blue}). Note that the 1.4\,mm SPT flux densities shown here are deboosted flux densities and not raw flux densities from which the original selection was made.
The top panel shows that the sample studied here populates the fainter part of the SPT-DSFG sample at both 1.4\,mm and 870\,$\mu$m, which is expected based on the selection method.
The redshift of the sources can be inferred from the 870\,\micron\ to 350\,\micron\ flux density ratio shown in the bottom panel (where sources with a higher 870\,\micron /350\,\micron\ flux density ratio typically are at higher redshifts).
This plot shows that the sample studied here is not expected to have a different redshift distribution compared to the full SPT DSFG sample.
Together with the \citet{weiss13} sample it is a representative sample of the full SPT-DSFG population.

The Cycle 1 ALMA 3\,mm spectral scans were carried out in 2013 July and 2013 December in the Cycle 1 early science compact array configuration. 
The observations were set up as spectral scans using five tunings to cover the 3\,mm atmospheric transmission window (See Figure \ref{Fig:CO-coverage}). 
Each tuning consists of two 3.75\,GHz wide sidebands covered by two 1.875\,GHz spectral windows in the ALMA correlator, which in total gives 7.5\,GHz coverage.
This setup spans $84.2$$-$$114.9$\,GHz, where the range $96.1$$-$$103.0$\,GHz is covered twice. 
Over this frequency range the FWHM of ALMA's primary beam is 61$''$$-$45$''$. 
The observations were not carried out for all tunings at the same time and the tunings were therefore observed with a different number of antennas ranging from 28 to 40, which resulted in typical synthesized beams of $3.7''\times2.4''$ to $3.0''\times1.8''$ from the low to high-frequency ends of the band.
The sources were observed for 120\,seconds each, in each tuning, which amounts to roughly 10\,minutes per source in total.
Typical system temperatures for the observations were $T_{\rm sys}$=60--90\,K (SSB). 
Flux calibration was performed on Uranus or Mars and passband and phase calibration were determined from nearby quasars. 
We used the CASA package to process the data. The cubes were created using natural weighting to optimize the sensitivity and constructed with a channel width of 19.5\,MHz (50--65\,kms$^{-1}$). 
The typical noise per channel is 1.5--2\,mJy\,beam$^{-1}$. 
The continuum images were also created and cleaned using natural weighting and were generated from the full bandwidth.
For these we have typical noise levels of 50\,$\mu$Jy\,beam$^{-1}$.

By scanning the 3\,mm window we are sensitive to CO lines between the CO(1--0) and CO(7--6) transition, which gives a redshift coverage of 0.0$<$$z$$<$0.4 and 1.0$<$$z$$<$8.6 with a narrow redshift desert at 1.74$<$$z$$<$2.00, see Figure \ref{Fig:CO-coverage}. For more details on the observation strategy and the spectral coverage of CO we refer to \citet{weiss13}.

\subsection{APEX/FLASH \cii\ follow-up} \label{sect:cii-obs}
For a subset of sources with only one line in the ALMA 3\,mm data (from either Cycle 0 or Cycle 1), we have performed APEX/FLASH \citep{klein14} observations in the 345\,GHz and 460\,GHz transmission window (see Table \ref{Tab:ALMApos} for a list of targets).
The data were obtained using Max Planck Society observing time in the period 2015 March to August. All observations were done in good weather conditions with an average precipitable water vapor of pwv$<$1.0\,mm, yielding typical system temperatures of $T_{\rm sys}$=240\,K. The observations were performed and the data processed in the same manner as described in \citet{gullberg15}.
Further details on the sources targeted in these observations can be found along with the \cii\ spectra in Appendix \ref{sect:supplementary_redshift_info} and \ref{sect:supplementary_singleline}.
\\

\subsection{APEX/SEPIA CO follow-up} \label{Sec:SEPIA}
For two sources (SPT0002-52 and SPT2349-50) we see a single bright line in the 3\,mm ALMA spectrum both with the most probable identification being CO(3--2).
We have obtained APEX/SEPIA 158 -- 211\,GHz \citep{billade12} observations confirming the redshift of these source by observing the CO(5--4) and CO(7--6) line for SPT0002-52 and SPT2349-50 respectively. The observations were carried out in 2015 September - November during ESO time (E-096.A-0939A-2015) under good weather conditions with an average precipitable water vapor pwv$<$1.0\,mm yielding typical system temperatures of $T_{\rm sys}$=150\,K. The data were reduced in the same manner as the APEX/FLASH \cii\ observations described above. Details on the sources along with the spectrum can be found in Appendix \ref{sect:supplementary_singleline}.
\\

\subsection{APEX/Z-Spec spectrum} \label{Sect:Zspec-obs}
For one source in the sample presented here (SPT0551-48), we used APEX/Z-Spec \citep[][]{naylor03,bradford09} to search for high-J CO lines in the frequency range 190 -- 310\,GHz and thereby identify the redshift of the source. The observations were obtained in November 2012 in good weather conditions. The reduction of the data was done in the same manner as described in \citet{bothwell13}. 
The resulting spectrum showed several lines identifying the redshift as $z$=2.5833(2) and it can be found  in Appendix \ref{sect:supplementary_singleline} along with a description of the source.
\\

\subsection{Photometry} \label{Sect:Photometry}
Here we provide an overview of the photometry measurements used to determine dust temperatures and assign probabilistic redshift estimates to the sources with single-line detections.

The ALMA 3\,mm flux densities were extracted as the peak flux density of the point sources on the cleaned continuum map and the error was determined as the rms in the area just around the source. 
For SPT2349-56, which is slightly spatially extended, we integrated over the entire source (see Section \ref{Sec:ALMA3mm}).

The SPT 1.4\,mm and 2.0\,mm flux densities were extracted and deboosted as described by \citet{mocanu13}.

With APEX/LABOCA we obtained 870\,$\mu$m flux densities in two Max Planck Institute observing programs in the period 2010 September - 2012 November.
The observations and data processing are described in \citet{greve12}.

The {\it Herschel}/SPIRE maps at 250\,$\mu$m, 350\,$\mu$m and 500\,$\mu$m were observed in two observing programs OT1\_jvieira\_4 and OT2\_jvieira\_5 in the period 2012 August -- 2013 March. 
The {\it Herschel}/SPIRE data consists of triple repetition maps, with coverage complete to a radius of 5 arcmin($'$) from the nominal SPT position. The maps were produced via the standard reduction pipeline HIPE v9.0.
The flux densities were extracted by fitting a Gaussian to the source and using the peak as the flux density. The flux densities have been corrected for pixelation as described in the SPIRE Observers Manual.
The noise was estimated by taking the RMS in the central few arcmins of the map which is then added in quadrature to the uncertainty due to pixelation.

The {\it Herschel}/PACS maps were obtained in the programs OT1\_jvieira\_4 and DDT\_mstrande\_1.
The data were recorded simultaneously at 100 and 160\,$\mu$m. Each scan comprises ten separate 3\,$'$ strips, each offset orthogonally by 4\,$''$. The two PACS maps were co-added, weighted by coverage.
The flux densities were extracted using apertures, with sizes fixed to 7\,$'$ for the 100\,$\mu$m map and for 10\,$'$ for the 160\,$\mu$m map.
The aperture sizes were determined based on Figure 17 and the aperture correction based on Table 15 in the PACS Photometer - Point-Source Flux Calibration document released from {\it Herschel}.\footnote{ 
http://herschel.esac.esa.int/twiki/pub/Public/\\
PacsCalibrationWeb/pacs\_bolo\_fluxcal\_report\_v1.pdf}
The uncertainty was obtained by random aperture photometry in the few central arcmins.

When fitting to the SEDs we have added in quadrature an absolute calibration uncertainty of 7\% for {\it Herschel}/PACS and 10\% for all other wavelengths.

\subsection{Ancillary spectroscopic observations}\label{Sect:Anc_spec}
In addition to the primary data presented here, we also make use of spectroscopic data taken at radio 
and optical wavelengths.

Simultaneously with the ALMA and APEX redshift confirmation observations, we carried out follow-up observations with Australia Telescope Compact Array (ATCA) targeting low-$J$ CO transitions. 
These data are presented in \citet{spilker14}, \citet{aravena13} and \citet{aravena15} and helped to secure some of the redshifts before the delivery of the ALMA data. 
Results from these observations are discussed in Section \ref{sect:add-spec} and included in Table\,\ref{Tab:Lines}.

Optical spectroscopy was performed for SPT2357-51 on the night of 2013 October 16 with the X-shooter echelon spectrograph \citep{vernet11} on the ESO VLT-UT2 (Kueyen) as part of program E-092.A-0503(A), with near-continuous spectroscopy from 0.3\,\micron\ to 2.48\,\micron\ with a 1$\farcs2$-wide and 11$''$-long slit.
Seeing conditions were $\sim$0.8$''$, taken at low average airmass of 1.2. The resolving power attained for our IR-channel observations was $R=5000$. The resolving power for the optical channel was $R=6700$.
We used the ESO pipeline \citep{modigliani10} to reduce our data. This pipeline applied spatial and spectral rectification to the spectra, and the data were flat-fielded and cosmic rays were identified and masked. The two dither positions were subtracted to remove the sky to first order, and the different echelle orders were combined together into a continuous spectrum (taking into account the variation in throughput with wavelength in different overlapping echelle orders) before spatially registering and combining the data taken at the two dither positions, and removing any residual sky background.

\section{Results} \label{sect:results}								 %
\subsection{Targeted ALMA 1\,mm observations} \label{Sect:alma-z-conf}

\begin{figure*}[t!]
	\centering
	\includegraphics[width=17.5cm,angle=0]{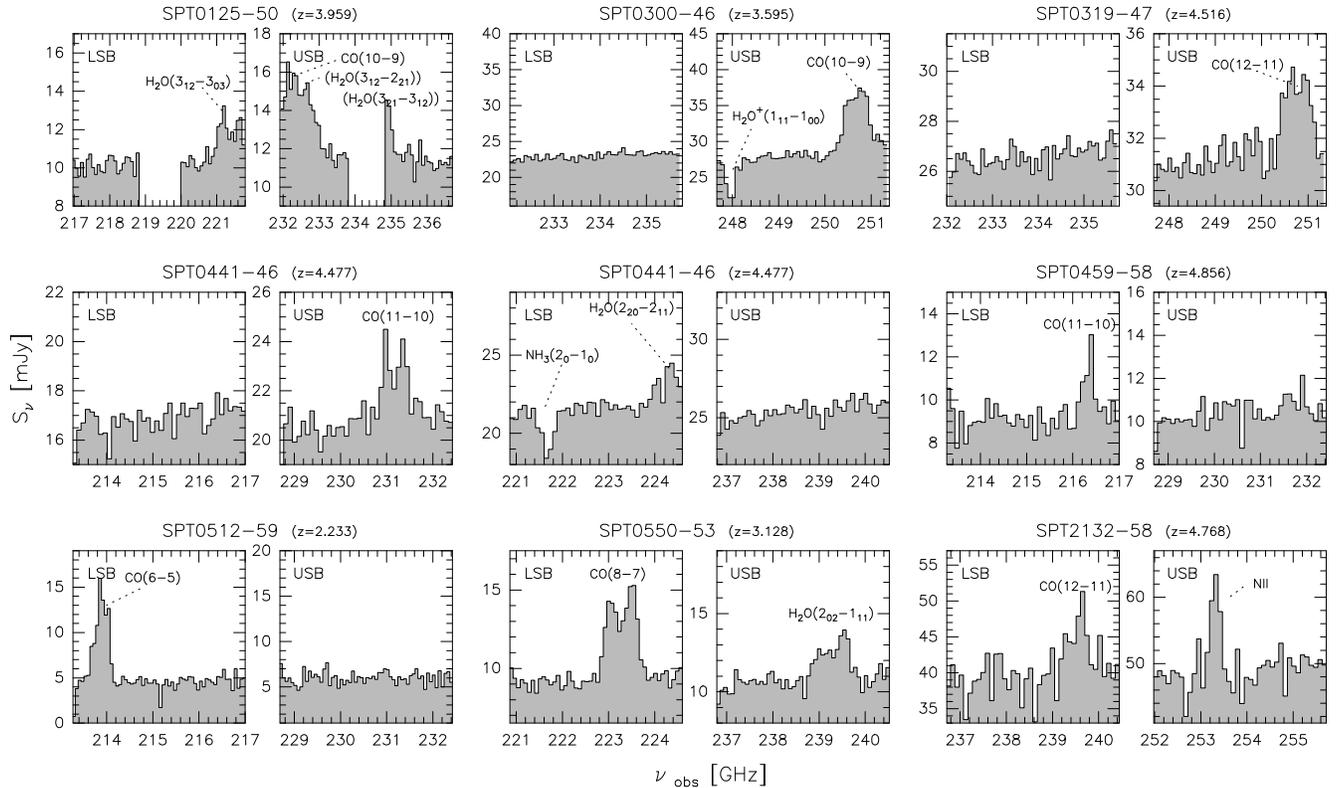}
	\caption{ALMA 1\,mm spectra for sources with redshifts based on a single submm emission line from \citet{weiss13} (see Section \ref{Sect:alma-z-conf}). For each source we show the LSB and USB spectra in the left and right panel, respectively. Each sideband has a total bandwidth of 3.9\,GHz.}
	\label{Fig:spectra_cy1} 
\end{figure*}

In the 1\,mm continuum images all sources but one are spatially unresolved and are detected with signal--to--noise ratios (SNRs) of 25--100. For the spatially resolved source (SPT0512-59, see Appendix \ref{sect:supplementary_redshift_info}), the brightest component is detected with a SNR of 9, and we extract the source spectrum from this component.
All 1\,mm spectra are shown in Figure\,\ref{Fig:spectra_cy1} (smoothed to lower velocity resolution for better visualization of the lines). 

We detect spectral line features in all sources, including emission lines from various CO transitions, \nii\ and several H$_2$O transitions and absorption lines from H$_2$O$^+$ and NH$_3$.  More details on the lines/transitions can be found in the description of the individual sources in Appendix \ref{sect:supplementary_redshift_info}.

The most important result from our ALMA 1\,mm observations, with respect to the source redshifts, is that they confirm the most probable redshifts as given in \citet{weiss13} for all except one source (see Table\, \ref{Tab:Lines}). The one exception had two almost equally likely redshift options and the source turned out to be at the slightly less likely redshift. 
As such, our 1\,mm follow-up observations demonstrate that reliable redshifts for DSFGs can be obtained when only a single line is detected in the 3\,mm redshift scan, provided that the dust continuum spectral energy distribution (SED) of the source is well sampled.

One of the sources included in our 1\,mm follow-up program (SPT0319-47) was presented as having no lines detected in its ALMA 3\,mm scan in \citet{weiss13}.  The 3\,mm spectrum, however, did show a very broad (${\rm FWHM}\sim1700$\kms), faint line feature at 104.4\,GHz. 
In our 1\,mm follow-up observations we now detect a highly significant line at 250.76\,GHz in this source. This detection identifies the 3\,mm and 1\,mm lines as CO(5--4) and CO(12--11) placing SPT0319-47 at $z$=4.516(4). This source was also detected in \cii\ with APEX/FLASH cementing the redshift (see Appendix \ref{sect:supplementary_redshift_info}).

\subsection{A misidentified redshift: The discovery and solution} \label{sect:add-spec}
SPT0551-50 was identified in \citet{weiss13} as a secure redshift at $z=2.1232(2)$ based on a single CO line detection (identified as CO(3--2)) in conjunction with a detection of the C{\scriptsize IV} line from the Very Large Telescope (VLT).
We afterwards failed to detect \cii\ with APEX \citep{gullberg15} and CO(1--0) with ATCA at this redshift.
In particular the non-detection of the ATCA line is very significant with $L'_{\rm CO(3-2)}$/$L'_{\rm CO(1-0)}$ $>$ 6 compared to $L'_{\rm CO(3-2)}$/$L'_{\rm CO(1-0)}$ $\sim$ 1.2 for sources with secure redshifts \citep{spilker14} where $L'$ is the line luminosity (in K\,km\,s$^{-1}$\,pc$^{2}$, see \citealt{solomon97}) and rules out the earlier redshift determination by \citet{weiss13}.
Based on this we re-visited the other redshift options. 
The favored option based on the dust continuum SED is $z$=3.163(3) with the line in the ALMA 3\,mm spectrum being CO(4--3). This redshift was confirmed by new \cii\ observations with APEX/FLASH.
The line observed with VLT and interpreted as C{\scriptsize IV} most likely originates from an unrelated lensing arc.
For a more detailed description on the source and a presentation of the above mentioned data see Appendix \ref{sect:supplementary_redshift_info}.

\subsection{New ALMA Cycle 1 3\,mm scans} \label{Sect:z-search}
\begin{figure*}[t!]
	\centering
	\includegraphics[width=17.5cm,angle=0]{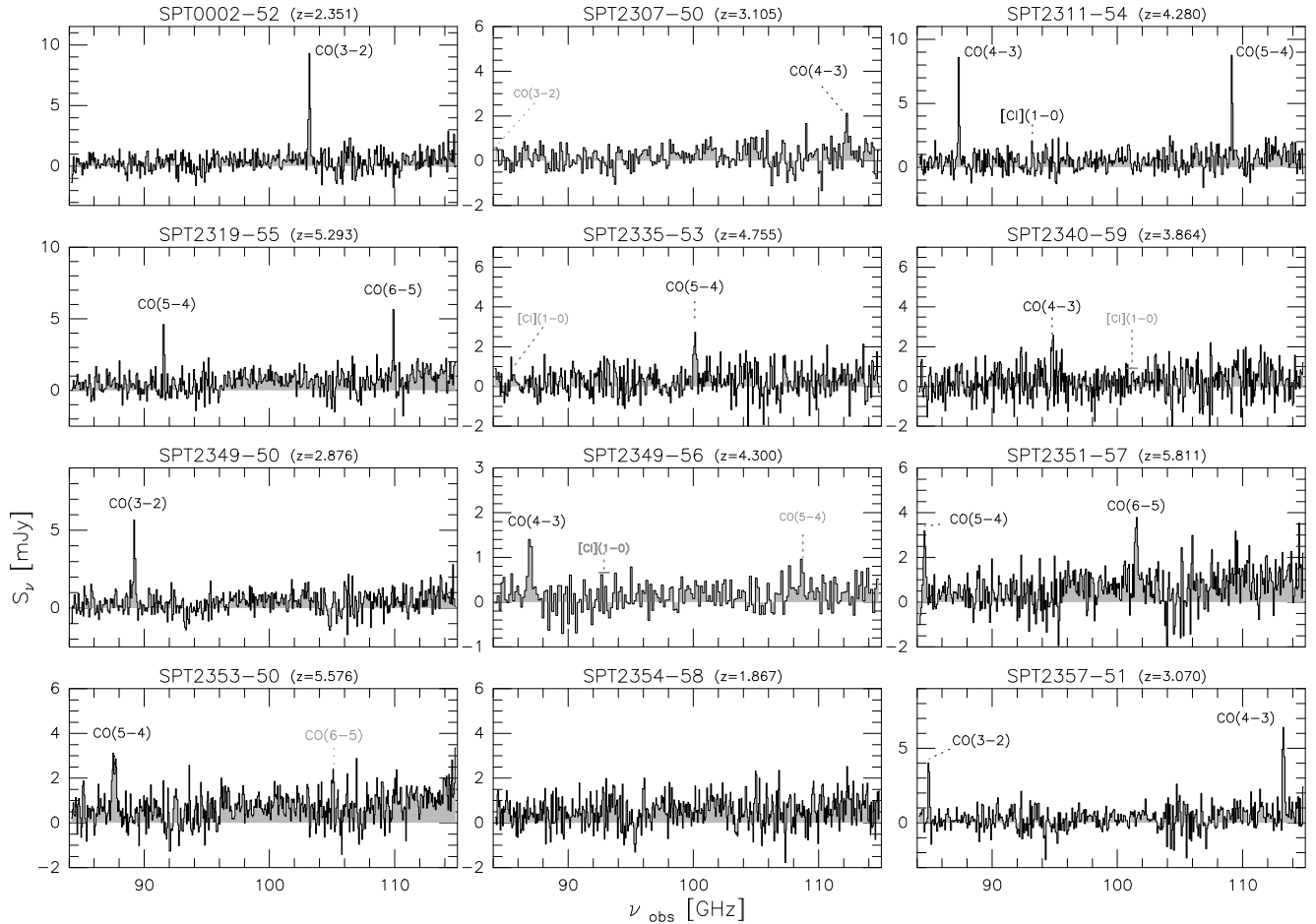}
	\caption{The 3\,mm ALMA spectra (spanning 84.2 -- 114.9\,GHz) of the 12 sources detected in continuum. Frequencies where we expect a line but we do not detect one are labelled in \emph{grey}. SPT2340-59 has multiple counterparts and the spectrum is extracted from counterpart B (see Appendix \ref{sect:supplementary_singleline}). SPT2349-56 likewise has multiple components and the spectrum shown here is a stack of the spectra extracted from components B and C.}
	\label{Fig:spectra} 
\end{figure*}

\subsubsection{Continuum results and morphology}
We detect the 3\,mm continuum in 12 out of 15 sources at SNRs of 5--18. 
For two sources the non-detection in ALMA was expected after a careful analysis of all photometric data (only available to us after ALMA's Cycle 1 deadline) showed that both sources are likely due to extended galactic foreground.\footnote{We submitted a source change request for these two sources to ALMA but the request was rejected.} 
These two sources have low SNRs in the SPT maps and are ignored in the following.
For the third source we would have expected a detection, but the non-detection is consistent with the overall SED of the source (see Appendix \ref{sect:SED_non-detection}).

Table \ref{Tab:ALMApos} lists the ALMA 3\,mm continuum position for the 12 detected sources. 
Their 3\,mm continuum flux densities are given in Appendix \ref{sect:supplementary_photometry}.

Ten sources appear as point sources, and two sources (SPT2340-59 and SPT2349-56) split into multiple components.  Multiple components can be explained in two ways: We either see multiple individual sources or multiple gravitationally lensed images of the lensed source. 
For SPT2340-59 we see two components (listed in Table \ref{Tab:ALMApos}, and named A and B respectively, see Appendix \ref{sect:supplementary_singleline} for the continuum image). We extract spectra at both positions, but only see a line in component A, which we then use in the further analysis.
For SPT2349-56 we see two components, one point source and one more extended component. 
For this source we take advantage of also having high resolution ALMA 870\,\micron\ imaging, which shows three counterparts. 
In this image the extended 3\,mm component breaks up into two point sources and we use the three 870\,\micron\ positions to define the three components A, B, and C listed in Table \ref{Tab:ALMApos} and shown in Appendix \ref{sect:supplementary_singleline}).
We see a hint of a line at the same frequency in all components (with small peak shifts), with component B and C showing stronger lines. It is not clear if the components belong to the same system (though the distance to component A suggests otherwise) or potentially only component B and C.
We use a stack of all three components in the further analysis.
This continuum morphology suggests that the two sources (SPT2340-59 and SPT2349-56) are made up of multiple distinct sources.

\subsubsection{Spectroscopy results}
The ALMA spectra of the 12 sources are presented in Figure \ref{Fig:spectra}. The spectrum for SPT2344-51 is not shown, since without the ALMA 3\,mm continuum detection, we do not know the source position with sufficient accuracy to be able to extract the spectrum. We verified, however, that the data cube for this source does not contain any strong lines. In total we see 16 lines in the 12 spectra, which we identify as $^{12}$CO and \ci\ emission lines.
We have marked the CO and \ci\ lines that we do not detect in \emph{grey}, where the horizontal line represents the expected flux density based on the SPT-DSFG line luminosities from \citet{spilker14}.

The lines are distributed over the sources in the following manner: \\\\
$\bullet$ Four sources show two or more lines, yielding an unambiguous redshift from the 3\,mm data alone (see Table \ref{Tab:Lines} top).\\
$\bullet$ Seven sources show a single line in the 3\,mm spectra. For three sources (SPT2335-53, SPT2349-56 and SPT2353-50) we have detected \cii\ with APEX/FLASH,  and for two sources (SPT0002-52 and SPT2349-50) we have detected CO using APEX/SEPIA. These additional lines secure the redshifts of the five sources. 
For the remaining two sources we use the dust SEDs to obtain the line identification and the redshift (see Section \ref{Sect:one-line}).\\
$\bullet$ One source shows no lines. For this source we find an absorption line in our 870\,\micron\ high resolution ALMA imaging cube determining the redshift (see Section \ref{Sect:no-lines}). \\

\begin{deluxetable*}{p{1.6cm} c >{\centering\arraybackslash}p{1.3cm} p{1.2cm} >{\centering\arraybackslash}p{0.7cm} p{4.7cm} p{5.4cm} }
\tabletypesize{\footnotesize}
\tablecaption{Redshifts and line identifications\label{Tab:Lines}}
\tablewidth{0pc}
\tablehead{
\colhead{Source} & \colhead{case} &\colhead{$z$} & \colhead{$T_{\rm dust}$} & $\lambda_{peak}$$\star$ & \colhead{lines from 3\,mm scans \textdagger} &  \colhead{new lines \& comments} \\
 & & & \colhead{[K]} & [\micron] & & 
}
\startdata
SPT2354-58 &	I	& {1.867(1)}		& 42.8$\pm$1.9	& 87		& no 3\,mm line				& $z_{phot} = 1.2 \pm 0.3$, OH$^+$ from ALMA	\\
SPT0452-50 &	II	&  {\bf 2.0104(2)}   	& 22.0$\pm$0.9        	& 145	& CO(3--2)\tablenotemark{a}	& CO(1--0)\tablenotemark{g} from ATCA\\ 
SPT0512-59 &	II	& {\bf 2.2331(2)}	& 32.7$\pm$1.4	& 106	& CO(3--2)\tablenotemark{a}	& CO(6--5) from ALMA; \cii \tablenotemark{d} from SPIRE FTS		\\
SPT0002-52 &	I	& {\bf 2.3513(4)}	& 42.3$\pm$2.1	& 88		& CO(3--2)				& CO(5--4) from APEX \\
SPT0125-47 &	II	&  {\bf 2.51480(7)}   	& 38.6$\pm$1.6       	& 93		& CO(3--2)\tablenotemark{a} \& CO(1--0)\tablenotemark{a} & \\
SPT0551-48 &	I	& {\bf 2.5833(2)}	& 38.6$\pm$1.9	& 93		& CO(7--6), CO(8--7) \& \ci (2--1) & lines from Z-Spec; CO(1--0)\tablenotemark{e} from ATCA; no ALMA data. 	\\
SPT2332-53 &	II	& {\bf 2.7256(2)}      	&47.4$\pm$2.8		& 81		& CO(7--6)\tablenotemark{a}, Ly$\alpha$\tablenotemark{a} \& C{\tiny IV}\,${\rm 1549\,\AA}$\tablenotemark{a}  & lines from Z-Spec; CO(1--0)\tablenotemark{e} from ATCA; no ALMA data \\ 
SPT2134-50 &	II	&  {\bf 2.7799(2)}   	&39.0$\pm$1.6        	& 93		& CO(3--2)\tablenotemark{a}, CO(7--6)\tablenotemark{a} \& CO(8--7)\tablenotemark{a}                  	& \\
SPT0538-50 &	II	& {\bf 2.7855(1)}   	&36.5$\pm$1.4		& 97		& CO(7--6)\tablenotemark{a}, CO(8--7)\tablenotemark{a} \& Si\,{\tiny IV}\,${\rm 1400\,\AA}$\tablenotemark{a}  
    & lines from Z-Spec; CO(1--0)\tablenotemark{e} and CO(3--2)\tablenotemark{f} from ATCA; no ALMA data \\ 
SPT2349-50 &	I	& {\bf 2.877(1)}		& 37.9$\pm$1.6	& 95		& CO(3--2)				& CO(7--6) from SEPIA	\\
SPT2357-51 &	I	& {\bf 3.0703(6)}	& 37.2$\pm$1.2	& 96 		& CO(3--2) \& CO(4--3)		& Lyman-$\alpha$ and OII$_{3727 \AA}$ from VLT/X-shooter\\ 
SPT0103-45 &	II	&  {\bf 3.0917(3)}   	& 32.3$\pm$1.2   	& 107	& CO(3--2)\tablenotemark{a} \& CO(4--3)\tablenotemark{a}           
    & \\
SPT2307-50 &	I	& {3.108(1)}		& 35.8$\pm$3.3	& 99		& CO(4--3)				& $z_{phot} = 3.4 \pm 0.9$	\\
SPT0550-53 &	II	& {\bf 3.1280(7)}	& 33.2$\pm$1.9	& 104	& CO(4--3)\tablenotemark{a}	
    & CO(8--7) from ALMA; \cii \tablenotemark{d} from APEX 						\\
SPT0551-50 &	II	&  {\bf 3.164(1)}   	&37.4$\pm$1.4        	& 96		& CO(4--3)\tablenotemark{b}                      
    & \cii\ and CO(8--7) from APEX \\
SPT0529-54 & II	&  {\bf 3.3689(1)}   	&31.8$\pm$1.2        	& 108	& CO(4-3)\tablenotemark{a}, \ci (1--0)\tablenotemark{a} \& $^{13}$CO(4--3)\tablenotemark{a}	&\\
SPT0532-50 &	II	&  {\bf 3.3988(1)}   	&37.6$\pm$1.4        	& 95		& CO(4-3)\tablenotemark{a}, \ci (1--0)\tablenotemark{a} \& $^{13}$CO(4--3)\tablenotemark{a}	& \\
SPT0300-46 &	II	& {\bf 3.5954(7)}	& 38.6$\pm$1.6	& 93		& CO(4--3)\tablenotemark{a} \& \ci (1--0)\tablenotemark{a} 
    & CO(10--9) from ALMA; \cii \tablenotemark{d} from APEX 	\\
SPT2147-50 &	II	&  {\bf 3.7602(3)}   	&40.2$\pm$1.6        	& 91		& CO(4--3)\tablenotemark{a} \& \ci (1--0)\tablenotemark{a}	& \\
SPT2340-59 &	I	& {3.864(1)}		& 40.2$\pm$1.9	& 91		& CO(4--3)		& $z_{phot} = 3.6 \pm 0.6$	\\
SPT0125-50 &	II	& {\bf 3.959(3)}		& 43.7$\pm$2.3	& 86		& CO(4--3)\tablenotemark{a} \& \ci (1--0)\tablenotemark{a}
    & CO(10--9) and H$_2$O\tablenotemark{c} abs line from ALMA  		\\
SPT0418-47 &	II	&  {\bf 4.2248(7)}   	& 45.3$\pm$2.3       	& 83		& CO(4--3)\tablenotemark{a} \& CO(5--4)\tablenotemark{a}	& \\
SPT0113-46 &	II	&  {\bf 4.2328(5)}   	& 31.3$\pm$1.4  	& 108	& CO(4--3)\tablenotemark{a}, \ci (1-0)\tablenotemark{a} \& CO(5--4)\tablenotemark{a}	& \\
SPT2311-54 &	I	& {\bf 4.2795(4)}	& 47.7$\pm$2.8	& 80		& CO(4--3), \ci (1--0) \& CO(5--4)	&					\\
SPT0345-47 &	II	&  {\bf 4.2958(2)}   	& 50.2$\pm$2.8       	& 78		& CO(4--3)\tablenotemark{a} \& CO(5--4)\tablenotemark{a}           
    & \\
SPT2349-56 &	I	& {\bf 4.304(2)}		& 46.7$\pm$2.8	& 82		& CO(4--3)					& \cii\ from APEX/FLASH \\	
SPT2103-60 &	II	&  {\bf 4.4357(6)}   	&37.4$\pm$1.6        	& 95		& CO(4-3)\tablenotemark{a} \& CO(5-4)\tablenotemark{a}		& \\   
SPT0441-46 & II	&  {\bf 4.4771(6)}   	& 38.1$\pm$1.9       	& 94		& \ci (1--0)\tablenotemark{a}, CO(5--4)\tablenotemark{a} \&  \cii \tablenotemark{a}
    & CO(11--10) \& NH$_3$ from ALMA\\
SPT0319-47 &	II	& {\bf 4.510(4)}		& $39.9 \pm$2.1	& 91		& CO(5--4)
    & CO(12--11) from ALMA; \cii\ from APEX			\\
SPT2146-55 &	II	&  {\bf 4.5672(2)}   	&37.4$\pm$2.1        	& 95		& \ci (1--0)\tablenotemark{a} \& CO(5--4)\tablenotemark{a}	& \\
SPT2335-53 &	I	& \bf{4.757(2)}		& 57.0$\pm$4.2	& 71		& CO(4--3)					& \cii\ from APEX 	\\  
SPT2132-58 &	II	&  {\bf 4.7677(2)}    	&37.9$\pm$1.9        	& 94		& CO(5--4)\tablenotemark{a} \& \cii \tablenotemark{a}     
    & CO(12--11) and \nii\ from ALMA \\
SPT0459-59 &	II	&  {\bf 4.7993(5)}   	&38.1$\pm$1.9        	& 94		& \ci (1--0)\tablenotemark{a} \& CO(5--4)\tablenotemark{a}	& \\
SPT0459-58 &	II	& {\bf 4.856(4)} 		& 41.6$\pm$1.9	& 88		& CO(5--4)\tablenotemark{a}
    & CO(11--10) from ALMA				\\
SPT2319-55 &	I	& {\bf 5.2929(5)}	& 42.1$\pm$2.1	& 87		& CO(5--4) \& CO(6--5)			&				\\
SPT2353-50 &	I	& {\bf 5.576(3)}		& 46.3$\pm$2.3	& 82 		& CO(5-4)	 					& \cii\ from APEX \\
SPT0346-52 &	II	&  {\bf 5.6559(4)}   	& 50.5$\pm$2.3       	& 77		& CO(5--4)\tablenotemark{a}, CO(6--5)\tablenotemark{a}, H$_2$O\tablenotemark{a} \& H$_2$O$^+$\tablenotemark{a}	& \\
SPT0243-49 &	II	&  {\bf 5.699(1)}   	& 32.7$\pm$1.6       	& 103	& CO(5--4)\tablenotemark{a} \& CO(6--5)\tablenotemark{a}	& \\
SPT2351-57 &	I	& {\bf 5.811(2)}		& 53.5$\pm$2.8	& 74		& CO(5--4) \& CO(6--5)		\\
SPT2344-51 &	III	&				&				& 		& no lines 				& $z_{phot} = 3.5 \pm 0.7$	\\
SPT0128-51 &	III	&				&				& 		& no lines\tablenotemark{a}	& $z_{phot} = 3.6 \pm 0.9$	\\
SPT0457-49 &	III	&				&				& 		& no lines\tablenotemark{a} 	& $z_{phot} = 3.4 \pm 0.6$	\\
\enddata
\tablecomments{The parenthesis at the end of the redshift gives the uncertainty on the last digit presented. The numbers in the column named 'case' refers to the following cases: {\bf I} New redshifts presented in this work. The unbolded redshifts show the single line redshifts {\bf II} Sources presented in \citet{weiss13}. Comments in the right column indicates observations added since then; {\bf III} Sources showing no lines.\\
\textdagger\  This column shows the lines from the 3\,mm line scan from this work and lines presented in \citet{weiss13}\\
$\star$ The rest frame SED peak wavelength.}
\tablenotetext{a}{Published by \citet{weiss13}}
\tablenotetext{b}{Published by \citet{weiss13} as CO(3--2)}
\tablenotetext{c}{Published by \citet{spilker14}}
\tablenotetext{d}{Published by \citet{gullberg15}}
\tablenotetext{e}{Published by \citet{aravena13}}
\tablenotetext{f}{Published by \citet{spilker15}}
\tablenotetext{g}{Published by \citet{aravena15}}
\end{deluxetable*}

\subsubsection{Sources with one detected line}  \label{Sect:one-line}
For the three sources where only a single line is detected in the 3\,mm spectrum, the first step in determining their redshift is to identify the possible line identifications. 
We work under the assumption that the line is from a transition in CO. For an overview of which lines are detectable in our 3~mm spectral scans from sources at a given redshift, see Figure \ref{Fig:CO-coverage} and \citet{spilker14}.
The most likely line identifications are either CO(2--1) or CO(3--2), as these lines appear in the observed band without any other lines present for a large redshift interval (1.0$<$$z$$<$3.0 with a narrow redshift desert at 1.7$<$$z$$<$2.0). At most redshifts CO(4--3) and CO(5--4) come with a \ci\ line in the observing band, and will only appear as single lines in very small redshift intervals or when the fainter \ci\ line remains undetected.   
We rule out $J$=6-5 and higher transitions as they will always come with another CO line in the observed band.

We use photometric data to determine the most likely redshift option for each source.  We do this as in \citet{weiss13}; the method is briefly described here. 
By fitting modified black body laws to the thermal dust emission of the sources, we can find the photometric redshift 
by assuming a dust temperature distribution for our sources; similarly, we can find the implied dust temperature
for a specific redshift option.
The thermal dust emission of the SPT-DSFGs is sampled by the following photometric observations: 3\,mm ALMA; 2 and 1.4\,mm SPT; 870\,$\mu$m APEX/LABOCA; 500, 350 and 250\,$\mu$m {\it Herschel}/SPIRE; and for most sources 160 and 100\,$\mu$m {\it Herschel}/PACS. This data is described in Section \ref{Sect:Photometry} and the flux densities are given in Appendix \ref{sect:supplementary_photometry}. 

\begin{figure}[t]
	\centering
	\includegraphics[viewport= 3 10 432 336, clip=true,width=8.0cm,angle=0]{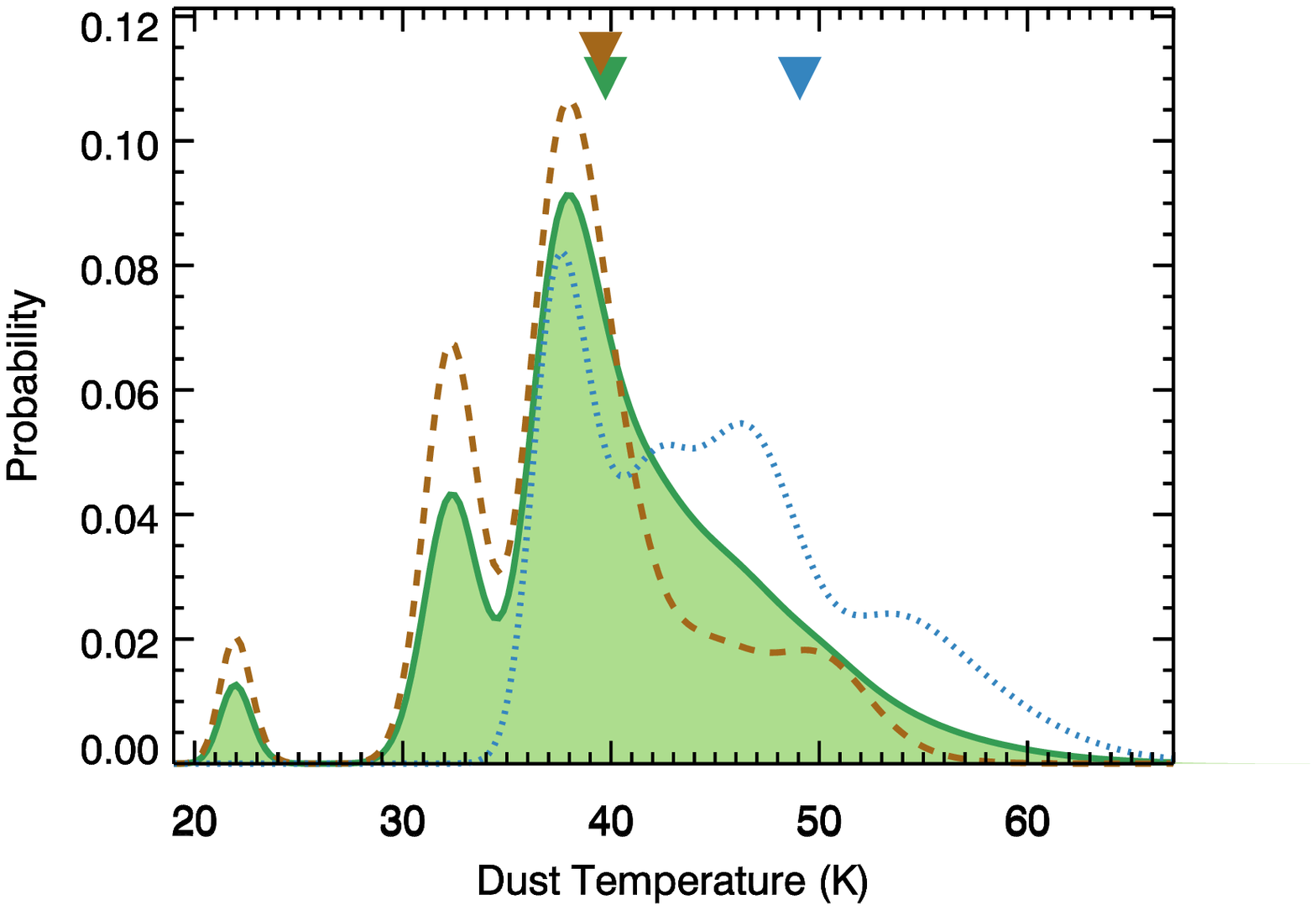}
	\caption{
  The probability distribution of the dust temperature for all 35 sources in the SPT-DSFG sample with unambiguous redshifts (\emph{green}). Overlaid is the dust temperature distribution for all sources observed in ALMA Cycle 0 \citep{weiss13} for which we have unambiguous redshifts (\emph{brown}) and for all sources from ALMA Cycle 1 with unambiguous redshifts (\emph{blue}). The triangles in the top of the plot show the median of the distributions.}
	\label{Fig:tdust} 
\end{figure}

To fit the SEDs we use the method described in \citet{greve12}, with a spectral slope of $\beta$=2, the optically thin/thick transition wavelength of 100\,$\mu$m, and taking the cosmic microwave background (CMB) into account. 
We ignore data shortwards of restframe 50\,$\mu$m in order to only fit the cold component of the thermal dust emission, as a single temperature SED can usually not describe all the thermal dust emission. 
The free parameters for the fit are dust temperature ($T_{\rm dust}$) and dust mass. 
Due to the degeneracy between dust temperature and redshift we have to assume a dust temperature to find the redshift. 
We investigate the dust temperature of our sample by fitting SEDs to all sources with a secure redshift and create a probability distribution for each source. 
We add these to create a probability distribution for the dust temperature for the full sample of sources with unambiguous redshifts. 
The probability distribution of the dust temperature is shown in Figure \ref{Fig:tdust} (\emph{green}), with the median dust temperature of $T_{\rm dust}$=39$\pm$10\,K indicated by the triangle. In this plot we have also plotted the temperature distribution for all sources from \citet{weiss13} with an unambiguous redshift (\emph{brown}) and for all sources in this work with an unambiguous redshift (\emph{blue}). In Table \ref{Tab:Lines} we show the dust temperature of each source along with their rest frame SED peak wavelength.

For each source, we create a photometric redshift probability distribution by randomly sampling $10^3$ temperatures from the $T_{\rm dust}$ distribution obtained from all sources with unambiguous redshifts and using this dust temperature in the SED fit for the source.
The peak of this distribution is then used as the photometric redshift with its errors reflected by the 1\,$\sigma$ confidence interval around the peak. This produces asymmetric errors but here we choose the larger of the two to be conservative.

\begin{figure*}[ht]
	\centering
	\includegraphics[viewport= 3 7 1060 336, clip=true, width=18.0cm,angle=0]{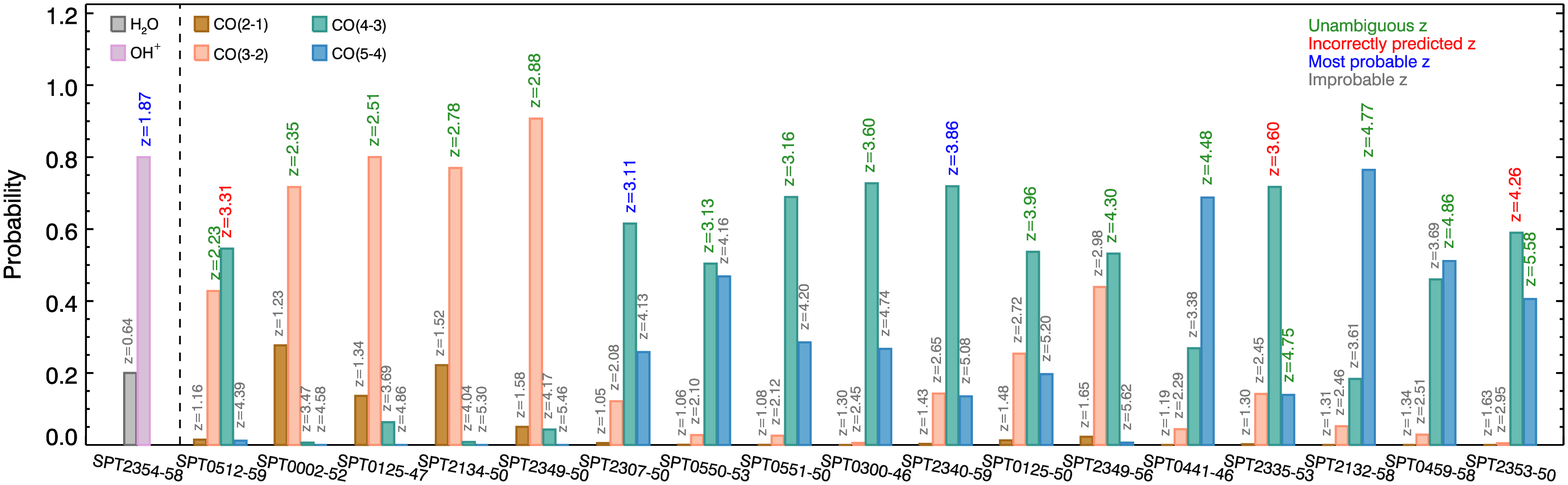}
	\caption{
   Histograms showing the probability of the redshift options for each source with a single line detection in the ALMA 3\,mm spectrum based on the dust temperature distribution shown in Figure \ref{Fig:tdust}.  To the far left we show a similar analysis for the line identifications for the absorption line in the 870\,\micron\ data for SPT2354-58.
   The sources are sorted by redshift (spectroscopic if available and otherwise most probable). The bars represent the possible line identification. Sources where the most probable redshift is identical to the spectroscopic redshift are highlighted in  \emph{green}; sources where the most probable redshift is not the spectroscopic redshift are highlighted in \emph{red}. Sources for which we do not yet have a confirmed spectroscopic redshift with at least two lines are highlighted in \emph{blue}.}
	\label{Fig:zprob}
\end{figure*}

We use this technique to test the redshift options for all four sources with a single line identification (see sources with redshifts given in \emph{blue} in Figure \ref{Fig:zprob}).
We calculate the probability of each redshift option by reading off the probability of each option from the photometric redshift probability distribution and normalizing the total probability to unity.

In Figure \ref{Fig:zprob} we show all sources with a single line in their ALMA 3\,mm spectrum from both \citet{weiss13} and this work.
We use the current dust temperature distribution (\emph{green} distribution in Figure \ref{Fig:tdust}) to calculate the probabilities for the line identifications for all the sources in the Figure. This means that this is not the same prediction as was made for the source before its redshift was found but it serves to show how well the current method predicts redshifts.
We show the spectroscopic redshift of each source in \emph{green} and where the prediction does not correspond to the spectroscopic redshift we have highlighted the redshift in \emph{red}. 
For the sources which do not have a spectroscopic redshift the most probable redshift is highlighted in \emph{blue}.
In a sample of 15 sources with a single line in their ALMA spectrum we correctly predict the redshift for 12 sources ($80$\% success rate). 
This seems to be a reliable but not perfect method so to be certain of the redshifts presented we continue our observing campaigns to obtain an extra line.
All sources from \citet{weiss13} which previously only had a single line observed now have a second line observed and thereby have secure redshifts.
The redshifts for the sources with single line detections are listed unbolded in Table \ref{Tab:Lines}. \\

In two of the cases where we find one line in the ALMA spectrum we would expect to detect a second CO line given their most probable or confirmed redshift.
For SPT2353-50 the ALMA 3\,mm line is found to be CO(5--4) based on the detection of \cii\ but we do not detect the CO(6--5) line that is expected to also be in the spectrum. It may be associated with a SNR$\sim$1.5 feature in the spectrum at the position of the line. 
Using the stacked spectrum of the SPT-DSFGs \citep{spilker14} we calculate the line luminosity ratio $L'_{\text{CO(6--5)}}$/$L'_{\text{CO(5--4)}}$$\sim$0.7 for the SPT-DSFGs presented in \citet{weiss13}.
The SNR$\sim$1.5 feature gives a line luminosity ratio of $L'_{\text{CO(6--5)}}$/$L'_{\text{CO(5--4)}}$$\sim$0.3.
For SPT2349-56, the ALMA 3\,mm line is found to be CO(4--3) based on the detection of \cii\ but we do not detect the CO(5--4) line, though it can be associated with a SNR$\sim$1.5 feature in the spectrum at the predicted frequency of the CO(5--4) line. From \citet{spilker14} the line luminosity ratio is  $L'_{\text{CO(5--4)}}$/$L'_{\text{CO(4--3)}}$$\sim$1.1. The SNR$\sim$1.5 feature gives a line luminosity ratio of $L'_{\text{CO(5--4)}}$/$L'_{\text{CO(4--3)}}$$\sim$0.7. In both cases the second line have a lower than expected line luminosity ratios but not inconsistent with typical line ratios found in high redshift sources \citep[e.g. see review by][]{carilli13}.

\subsubsection{Sources without ALMA line detections} \label{Sect:no-lines}

As mentioned above, we searched for emission lines in the ALMA data cube of SPT2344-51 (which remained undetected in the 3\,mm continuum data), but did not find evidence for any strong lines despite its photometric redshift of $z_{phot}$=3.5$\pm$0.7. Given the faintness of this source in the continuum, the most likely interpretation is that it is simply too faint to detect its CO lines at the sensitivity limit of our observations. This most likely also holds for the two remaining sources without line detections from \citet{weiss13} (SPT0128-51 \& SPT0457-49). Their 1.4\,mm and 870\,\mue\ continuum flux densities are comparable to those of SPT2344-51, which place them at the faint end of the SPT sources targeted with ALMA though we do detect these sources in continuum. 
For SPT0457-49 we searched the redshift desert with ATCA looking for CO(1--0) without success (see Appendix \ref{sect:supplementary_redshift_info}).
Without deeper data we cannot determine their redshift and we drop these three sources in the analysis of the redshift distribution.
A summary of the sources for which we do not detect any  lines is presented at the bottom of Table\,\ref{Tab:Lines}. 

The situation is different for SPT2354-58, which does not show indications for CO lines in the ALMA Cycle 1 3\,mm spectrum. Here the continuum flux densities (see Appendix\,\ref{sect:supplementary_photometry}) are such that we should have detected CO lines based on the line to continuum ratio of the SPT sources where we detect lines.
For this source we find an absorption line in our 870\,\micron\ high resolution imaging data cube. This absorption line has two line identifications that fall outside the redshift range probed by the 3\,mm redshift search: OH$^+$(1$_{22}$ -- 0$_{11}$) at $z$=1.867(1) (in the redshift desert) and H$_2$O($1_{10}-1_{01}$) at $z$=0.6431(3) (below our searched redshift range). 
There  could possibly be more molecules which may show up in absorption, but we limit the discussion to the most likely ones. 
OH$^+$(1$_{22}$ -- 0$_{11})$, unlike H$_2$O($1_{10}-1_{01})$, has been detected in the local ultra luminous galaxy Arp220 \citep{rangwala11}.
Furthermore, for the second option we should have seen CO(5--4) in the same cube but we did not. The first redshift option is also preferred by the photometry (see Figure \ref{Fig:zprob}) and it is thus the most likely redshift and we have added this source to our list of sources with single-line redshifts (see Table \ref{Tab:Lines}). The ALMA 870\,\micron\ spectra are shown along with a more detailed description of the source in Appendix \ref{sect:supplementary_singleline}.

\section{Discussion} \label{sect:discussion}						 %

\subsection{The redshift distribution}

Our sample is composed of 39 sources with reliable redshifts (three from APEX/Z-Spec and 36 from ALMA 3\,mm scans), meaning that they show at least one spectral line along with well-sampled photometry. 
This translates into a success rate for our ALMA 3\,mm scan technique of $>85\%$ (36 out of 41 targeted).
Two or more lines have been identified in 35 of the 39 sources ($\sim80\%$). For 18 sources the redshifts were identified directly from the ALMA 3\,mm spectrum, for 3 sources the redshifts were found using Z-Spec/APEX, and for the remaining 13 sources the redshift was secured with observations of a second molecular line from ALMA, APEX, ATCA or {\it Herschel}/SPIRE.

\begin{figure}[t]
	\centering
	\includegraphics[viewport=3 48 370 283, clip=true,width=8.0cm,angle=0]{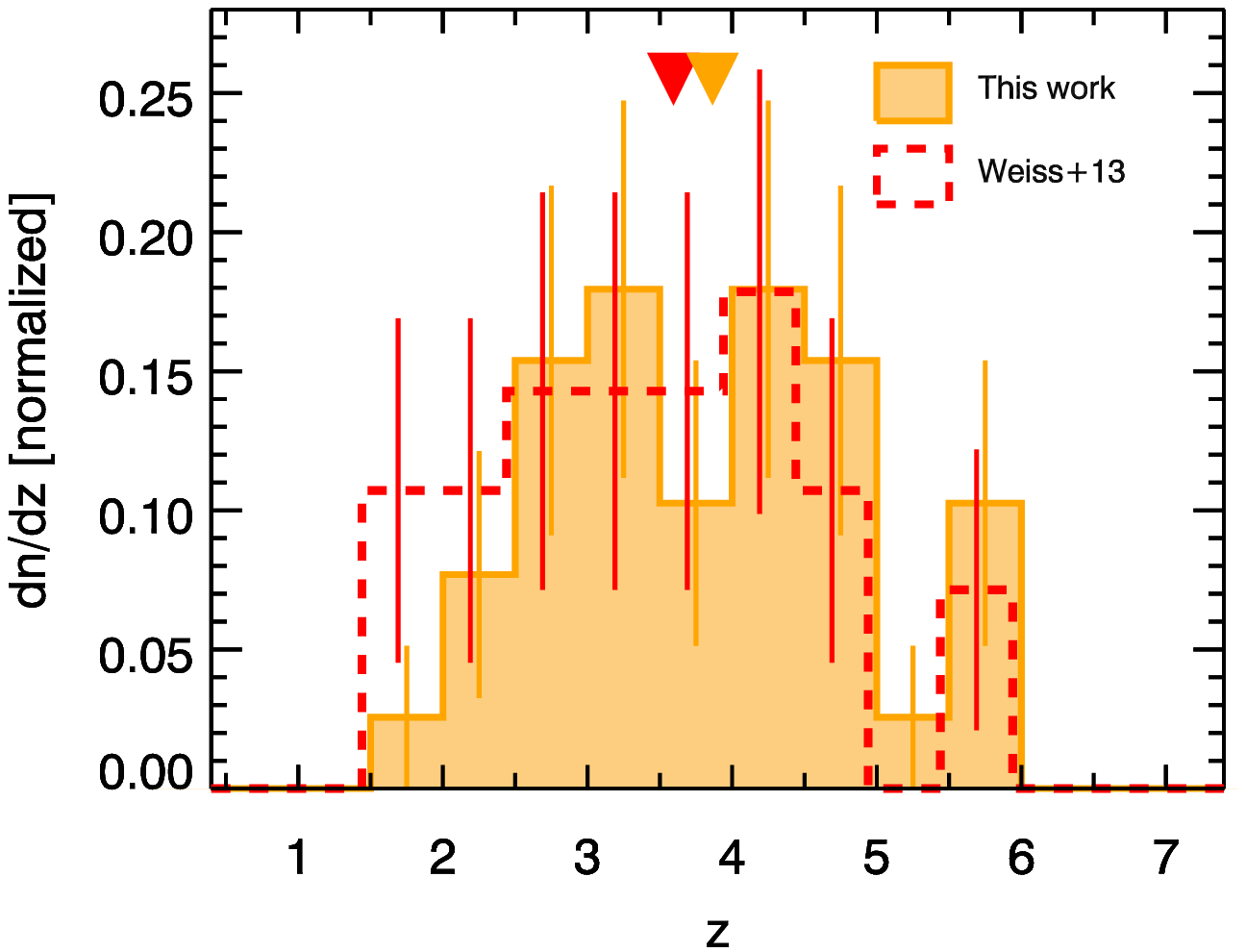}\\
	\includegraphics[viewport=3 48 370 283, clip=true,width=8.0cm,angle=0]{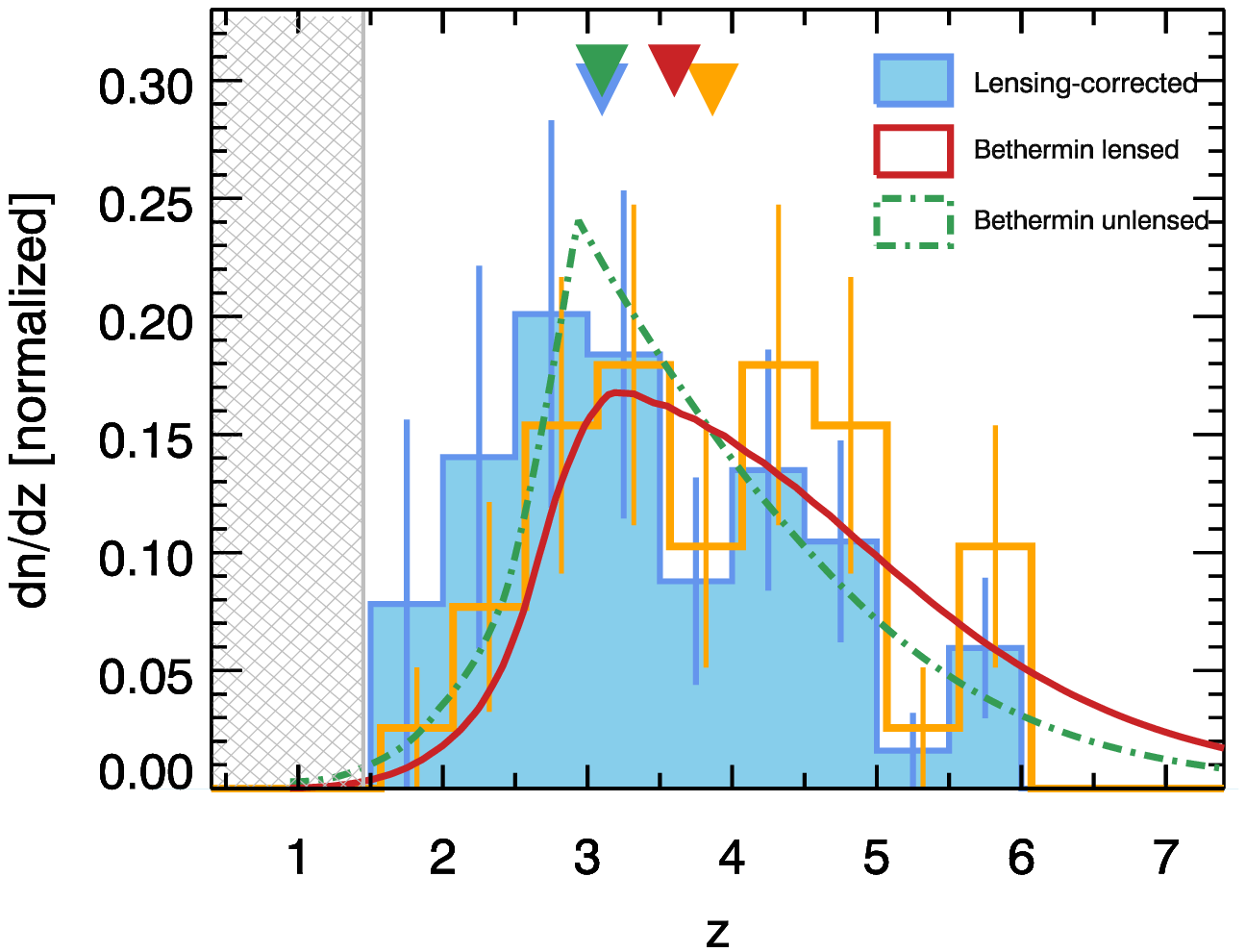}
	\includegraphics[viewport=3 0   370 283, clip=true, width=8.0cm,angle=0]{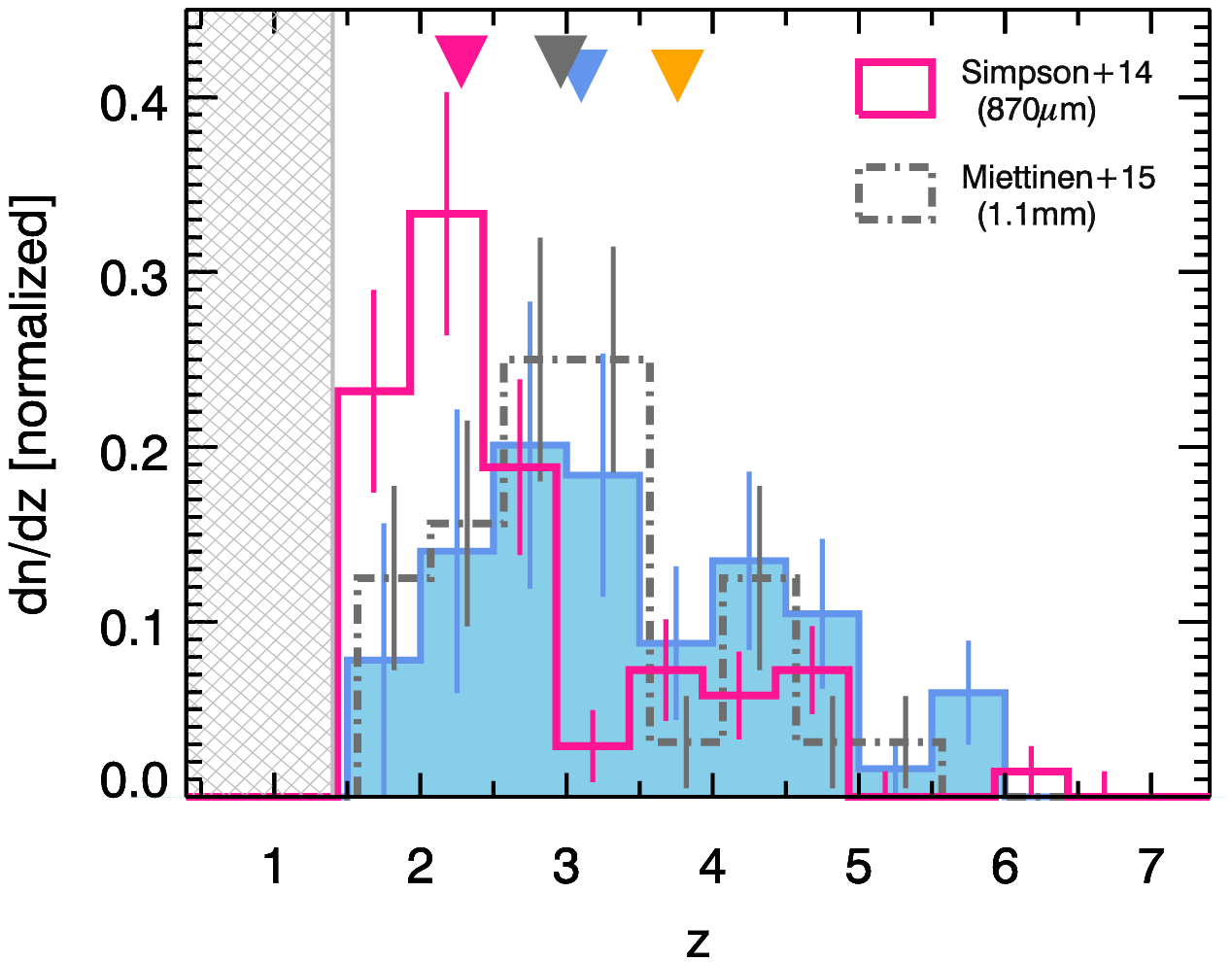}
	\caption{Redshift distributions for various samples described in Section \ref{sect:discussion}. The filled triangles in the top of the plots show the median redshifts of the samples with the corresponding colors.
{\it Top}: The \emph{orange} histogram shows the redshift distribution of the sources in our sample. It is compared to the original redshift distribution from \citet{weiss13} (\emph{red} dashed line).
{\it Middle}: The lensing-corrected redshift distribution of our sample (\emph{blue}) compared to the uncorrected redshift distribution (\emph{orange}, identical to the \emph{orange} shaded region in the top panel). The grey hatched region indicates the region we do not include in our analysis due to the low probability of finding lensed sources (see Figure \ref{Fig:lens-prob}). Also shown are model predictions from \citet{bethermin15} of a sample of lensed sources selected in the same way as the SPT-DSFGs (\emph{red}) and a sample of unlensed sources selected in the same way as the SPT-DSFGs (\emph{green}).
{\it Bottom}: The lensing-corrected redshift distribution compared to redshift distributions from \citet[][\emph{pink}]{simpson14} and \citet[][\emph{grey}]{miettinen15}, where sources below $z$$<$1.5 have been removed for a fair comparison.
} 
	\label{Fig:NZ}
\end{figure}

The redshift distribution of this sample is shown in \emph{orange} in Figure \ref{Fig:NZ} and listed in Table \ref{Tab:dndz}. The median redshift is $z$=3.9$\pm$0.4 (indicated by an \emph{orange} triangle above the distribution). 
The errors on the median were determined using a bootstrap method, where we randomly sampled 39 sources from the redshift distribution 1000 times and took the standard deviation of the median values.

The distribution is flat between $z$=2.5 and $z$=5.0 with a large fraction (75\%) of the sample at $z$$>$3.
We see no sources at $z$$<$1.5 as the probability of a source undergoing strong gravitational lensing drops significantly below a redshift of $z$$\sim$$2$.

In the top panel of Figure \ref{Fig:NZ} we overlay the distribution from \citet{weiss13} (dashed \emph{red} line).
As the sample in \citet{weiss13} was selected from 1300 square-degrees with S$_{1.4 {\rm mm}}$$ >$$ 20$ mJy, it is representative of the brightest sources from the SPT-DSFG sample. 
The difference in flux at various wavelengths between the two samples is shown in Figure \ref{Fig:sample}. Models from \citet{bethermin15} predict that the difference between these two samples based on the change in flux cut are negligible.
The only redshift bins in which we see a significant difference between this sample and that of \citet{weiss13}, are those in the range 1.5$<$$z$$<$2.5. 
In \citet{weiss13}, all three sources without line detections were placed in the 1.5$<$$z$$<$2.5 bin assuming they fell into the CO redshift desert (1.74$<$$z$$<$2.00). As discussed in Section \ref{Sect:no-lines}, we do not follow this approach, but ignore sources without detected lines.
Removing these sources from the distribution of \citet{weiss13} (except for SPT0319-47 which now enters with an unambiguous redshift) and correcting the one misidentified redshift (SPT0551-50, see Section \ref{sect:add-spec}) changes the median from $z$=3.6 to $z$=3.8, which is consistent with what we find in this work. A KolmogorovÐ Smirnov (K-S) test show that the probability that these two distributions originate from a common sample is  $p$$=$$0.81$.

\begin{deluxetable}{ccccccc}
	\centering
	\tablecaption{\label{Tab:dndz} Measured redshift distribution for SPT sources}
	\startdata
	\tableline
	\tableline
         $z$  & & N\tablenotemark{a} &&d$n$/d$z$ & &lens-cor\tablenotemark{b} d$n$/d$z$\\
	\tableline
        $1.5-2.0$ & \,\,	& 1 & \,\, 	& 0.03 $\pm$ 0.03 & \,\,\,	& 0.08 $\pm$ 0.08 \\
        $2.0-2.5$ &	& 3 &	& 0.08 $\pm$ 0.06 &		& 0.14 $\pm$ 0.08 \\
        $2.5-3.0$ &	& 6 &	& 0.15 $\pm$ 0.06 &		& 0.20 $\pm$ 0.08 \\
        $3.0-3.5$ &	& 7 &	& 0.18 $\pm$ 0.07 &		& 0.18 $\pm$ 0.07 \\
        $3.5-4.0$ &	& 4 &	& 0.10 $\pm$ 0.05 &		& 0.09 $\pm$ 0.04 \\
        $4.0-4.5$ &	& 7 &	& 0.18 $\pm$ 0.07 &		& 0.13 $\pm$ 0.05 \\
        $4.5-5.0$ &	& 6 &	& 0.15 $\pm$ 0.06 &		& 0.10 $\pm$ 0.04 \\
        $5.0-5.5$ &	& 1 &	& 0.03 $\pm$ 0.03 &		& 0.02 $\pm$ 0.02 \\
        $5.5-6.0$ &	& 4 &	& 0.10 $\pm$ 0.05 &		& 0.06 $\pm$ 0.03 \\
	\tableline
	\tablecomments{The numbers stated here are based on 39 sources, where all sources have at least one spectral line and well sampled photometry.}
	\tablenotetext{a}{Number of sources per bin.}
	\tablenotetext{b}{d$n$/d$z$ for lensing-corrected redshift distribution.}
\end{deluxetable}

\subsection{Selection effects}
This section describes the influence of our selection methods on the redshift distribution.
The two main effects come from our high flux cut that selects almost exclusively gravitationally lensed sources and our long selection wavelength.

As discussed in \citet{blain02}, \citet{dacunha13}, and \citet{staguhn14} the CMB could make cold DSFGs at high redshifts difficult to detect.
As SPT-DSFGs have a median dust temperature of $T_{\rm dust}$=39\,K, and are thus quite warm,  this effect only becomes relevant at very high redshifts ($z$$>$10).
For the sources presented in this work, the effect of the CMB is negligible.

In the two following sections we describe the two main selection effects, gravitational lensing and wavelength selection, which complicate a direct comparison between the  redshift distributions of DSFGs in the literature.

\subsubsection{Lensing effects and lensing-correction} \label{sect:lens-effect}
\begin{figure}[t]
	\centering
	\includegraphics[viewport=7 2 390 250, clip=true,width=8.0cm,angle=0]{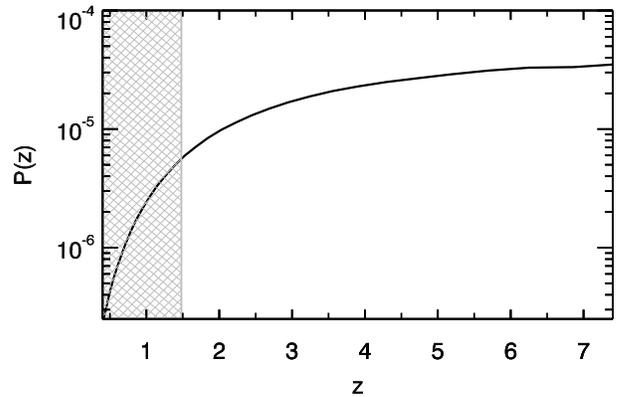}\\
	\caption{The assumed probability of gravitational lensing as a function of redshift for a source magnification of $\mu$=10 \citep{hezaveh11}. Because of the drastically falling probability of lensing below $z$$<$1.5 (\emph{grey} hatched region) we do not conclude anything about the redshift distributions in this range.} 
	\label{Fig:lens-prob}
\end{figure}
Based on models of the high-redshift DSFG population \citep[e.g.][]{baugh05,lacey10,bethermin12b,hayward13}, we would expect very few sources intrinsically bright enough to exceed our adopted flux  density threshold at 1.4\,mm 
($>$16\,mJy), and we thus expect the SPT-DSFG sample to consist almost solely of gravitationally lensed sources \citep{blain96,negrello07}.
This expectation was confirmed by ALMA 870\,$\mu$m high resolution observations showing that our sources resolve into arcs, multiple images, and Einstein rings, characteristic of lensed objects \citep[][]{vieira13, hezaveh13, spilker14}.
We have found a few examples where the source splits into multiple, ultra luminous galaxies (e.g., SPT2349-56; see Appendix \ref{sect:supplementary_singleline}) but these are rare in our sample.

Gravitational lensing is what enables us to study these high redshift sources in detail but also hampers a direct comparison of our results to unlensed samples.
Figure \ref{Fig:lens-prob}, based on the model presented in  \citet{hezaveh11}, shows the probability of a source undergoing strong
gravitational lensing between a given source redshift and the observer. The probability of sources at 
$z$$\lesssim$1.5 undergoing strong lensing is heavily suppressed relative to sources at higher redshifts ($z$$>$$4$), where the probability of lensing is flat. 
The lensing probability at $z$$\sim$2 is suppressed by a factor of three
compared to at high redshifts, while at $z$$\sim$3 this is reduced to a factor of less than two. All of these 
findings assume that DSFGs do not undergo a systematic size evolution with increasing redshift. 
Detailed discussions of this point can be found in \citet{weiss13}, \citet{bethermin15}, \citet{ikarashi15}, and \citet{simpson15}, in which it is concluded that there is no evidence for size evolution with redshift, although there are not enough measurements at high redshift ($z$$>$$4$) to  exclude the possibility of size evolution entirely.

To compare our sample to other samples from the literature, we correct our redshift distribution for the effect of gravitational lensing. 
We do this by dividing the redshift distribution by the probability for strong gravitational lensing as a function of redshift using the average magnification of our sample of $\mu$$\sim$10 (see Figure \ref{Fig:lens-prob}).
This yields the \emph{blue} redshift distribution shown in the middle panel of Figure \ref{Fig:NZ}, and
the median of our distribution decreases from $z$$=$$3.9\pm0.4$ to $z$$=$$3.1\pm0.3$ after the lensing-correction. 
To be able to calculate the error on the median of the lensing-corrected sample we randomly sample 39 sources 1000 times from the lensing-corrected redshift distribution. For each sample we find the standard deviation of the median and we use the mean of these as the error.
A K-S comparison of the observed and lensing-corrected distributions gives a value of $p$$=$$0.23$. 
In other words, gravitational lensing does not appear to have a statistically significant impact on our measurement of the redshift distribution of DFSGs. 

We use a single magnification to produce our lensing-corrected redshift distribution, which is a simplified approximation. 
The observed range of magnifications from the SPT DSFGs is $1$$<$$\mu$$<$$33$ with $<$$\mu$$>$$\approx$$9$ \citep[see][and Spilker et al. {\it in prep.}]{hezaveh13}. 
The relative shape of the lensing probability kernel (Figure \ref{Fig:lens-prob}) for different magnifications is identical, but offset vertically for higher or lower magnifications. 
It is not obvious, \textit{a priori}, that using a single magnification factor to lensing-correct our redshift distribution is a valid assumption.
To test this assumption, we compare our lensed and lensing-corrected redshift distributions to a lensed and unlensed model population from \citet{bethermin15} in Figure \ref{Fig:NZ} (middle panel). 
As discussed above, we have ignored sources in the model with $z$$<$1.5. 
The lensed model population is created using the same selection criterion as the SPT-DSFGs i.e. $S_{1.4mm}$$>$16\,mJy (\textit{dark red}) with an analytic model which includes gravitational lensing as in \citet{hezaveh11}. 
The model agrees well with our observed redshift distribution (Sum of squared residuals, weighted by the inverse square of the errors $\chi^2$=8.9 over 9 bins, median redshift = 3.6/3.9 for model/observed distributions).
The good agreement also holds if we use a slightly higher flux cut of $S_{1.4mm}$$>$ 25\,mJy as in \citet{weiss13}.
The unlensed model population is selected from the same model by ``demagnifying'' the SPT DSFG flux cut by a factor of $\mu$$=$$10$, i.e. $S_{1.4mm}$$>$1.6\,mJy (\textit{green}). 
With such a low flux cut, the number counts of unlensed sources completely dominates the lensed source counts. 
The model prediction for this sample has a median of $z$=3.1, in excellent agreement with our lensing-corrected redshift distribution, with $\chi^2$=9.7 over 9 bins.
The good agreement between the redshift distributions indicates that our simple method of lensing-correcting using a single magnification is a satisfactory approximation.

\subsubsection{The influence of the selection wavelength} \label{sect:sel-wave}

The other major influence on our redshift distribution is the selection wavelength. 
As  discussed in  \citet{blain02}, \citet{zavala14}, and  \citet{casey14},  the source selection function of mm and submm surveys varies with redshift, which  affects the observed redshift distribution. In general, for surveys down to mJy depths, a longer wavelength selection will lead to a higher observed redshift distribution. 

In the bottom panel of Figure \ref{Fig:NZ}, we compare our lensing-corrected redshift distribution to redshift distributions from the literature selected at different wavelengths. To make the distributions comparable to ours, we have removed  sources with redshifts below $z$$<$1.5, because 
the probability of strong gravitational lensing as a function of source redshift strongly disfavors the presence of these sources in our SPT-DSFG sample (see Figure \ref{Fig:lens-prob}).
We focus on the redshift distributions published  since \citet{weiss13}. 
As these distributions were selected from a small area ($<$1 deg$^2$) on the sky using a lower flux cut and are therefore made up of mostly unlensed sources, we compare to our lensing-corrected redshift distribution.

Using a selection wavelength of 1.1\,mm, \citet{miettinen15} presented redshifts for 15 galaxies from the COSMOS field, discovered with JCMT/AzTEC and followed up with high-resolution PdBI imaging. 
They add these new 15 sources to the 1.1\,mm selected sources from \citet{smolcic12}, also found using JCMT/AzTEC, and updated the redshifts where better data was available.
The final sample consists of  30 sources selected at 1.1\,mm with a mix of photometric and spectroscopic redshifts (see Figure \ref{Fig:NZ}, \emph{grey}). 
(Note that the distribution looks slightly different from the one shown by \citet{miettinen15} as they use probability functions for their redshifts and we use the redshifts given in their Table 4.) 
The median of this distribution is $z$=3.0, similar to ours, with a K-S comparison probability of $p$$=$$0.29$ which means that these two distribution are likely to be from the same common distribution. 

\citet{simpson14} created a sample of 97 870\,\micron\-selected sources, using high resolution ALMA data to identify the counterparts. 
They present a photometric redshift distribution containing 77 sources (where a fraction has spectroscopic redshifts) from the ALESS catalogue \citep{hodge13, karim13}, which is a sample of ALMA 870\,\micron-confirmed sources from the ECDF-S. 
The redshift distribution is shown in the bottom panel of Figure \ref{Fig:NZ} (\emph{pink}).
The photometric redshifts are based on a combination of radio, submm and NIR-optical data, and only sources with four or more data points are considered. 
Their median photometric redshift of $z$=2.3 is consistent with what was found by \citet{chapman05}, though the redshift distribution of \citet{simpson14} shows an excess of high-redshift sources over the earlier work, which relied on radio-wavelength counterpart identification. 
Their distribution differs significantly from ours with a K-S comparison probability $p$$=$$0.03$. In the paper they present another 19 sources with less than four photometry points. The redshifts for these are not given, but we tried to add 19 sources randomly in the redshift range 2.5$<$$z$$<$6.0 which gives a K-S comparison probability of $p$$=$$0.04$. Both these values are below  $p$$<$0.05, indicating they are not drawn from the same distribution.

In addition to studying the effect of gravitational lensing on the redshift distribution of DSFGs, 
\citet{bethermin15}  studied how their model predicts the shape and median of redshift distributions for samples selected at different wavelengths. 
They found that the difference in redshift distributions seen in Figure \ref{Fig:NZ} can be reasonably explained by the wavelength selection. 
Both the distribution of \citet{simpson14} and \citet{miettinen15} follow these predictions,
although the distributions selected around 850\,\micron\ put some strain on the models, as they are peaking at slightly lower redshifts than predicted.
While there may remain some questions as to the redshift completeness and reliability of photometric redshifts in the two comparison samples, we interpret the selection wavelength as the main driver for the difference
in redshift distributions seen in the bottom panel of Figure \ref{Fig:NZ}.

\section{Summary and Conclusion} \label{sect:summary}				 %

We have used ALMA in Cycle 1 to determine spectroscopic redshifts for strongly lensed DSFGs selected from the SPT survey.
With this data, we confirmed the redshifts of six sources with single-line redshifts from Cycle 0 presented in \citet{weiss13} and performed a redshift search for 15 new sources. 

Observing in Band 6 for 8--20 minutes per source, we have measured mid- to high-$J$ CO lines to confirm previously reported single-line redshifts from \citet{weiss13}.  In addition to detecting lines originating from transitions in CO, we also detected \nii , H$_2$O, H$_2$O$^+$ and NH$_3$. The most probable redshifts (based on one line plus well sampled photometry) for all but one source were confirmed to be correct, demonstrating a robust method to estimate redshifts from a single line and a well measured dust temperature. This method will be useful for future blind surveys with ALMA. 

We sought redshift identification for 15 new sources selected from a 100 deg$^2$ field of the SPT survey with $S_{1.4mm}$$>$16\,mJy, by searching for emission lines in ALMA Band 3. We covered the frequency range 84.2--114.9\,GHz in five tunings of 2 minutes each, adding up to 10 minutes of integration per source.
Twelve of these sources are detected in continuum and their spectra are extracted.
In four sources, we find two or more lines and unambiguously determine the redshift. 
In seven sources we find one single line and calculate the most probable redshift for each of them using their dust temperature. 
For two of these sources we detect \cii\ with APEX/FLASH and for two sources we detect CO with APEX/SEPIA, securing their redshift.
In one source we do not see any lines in the 3\,mm ALMA spectrum, but we determine the redshift from an absorption line detected in our ALMA Cycle 0 870\,\micron\ high resolution imaging cube.

In total, we determine reliable redshifts for 12 sources targeted in our ALMA Cycle 1 3\,mm scans, present a redshift found using APEX/Z-Spec, and confirm six single-line redshifts from \citet{weiss13} with our targeted 1\,mm scans. Adding this to the already established redshifts of SPT-DSFGs gives a final sample of 39 sources with spectroscopic redshifts. The median of the sample is $z$$=$$3.9\pm0.4$ with a slightly lower mean of $\bar{z}$=3.7. 
Unlike redshift distributions selected at slightly shorter wavelengths, the SPT-DSFG redshift distribution is flat between $z$=2.5--5.0 with a large fraction (75\%) of the sample at $z$$>$3.


Assuming no size evolution with redshift, we lensing-correct the redshift distribution by taking into account the probability of gravitational lensing occurring as a function of redshift.
After correction for lensing, we recover the redshift distribution of DSFGs above $z$$>$1.5 and we find a median of $z$$=$$3.1\pm0.3$ for DSFGs selected at 1.4 mm. 
The redshift distribution and the lensing-corrected redshift distribution are consistent with the prediction made by the models of \citet{bethermin15}.

By comparing to redshift distributions from the literature, we show that the selection wavelength is an important variable to the shape of the redshift distribution. 
The long selection wavelength (1.4\,mm) of the SPT DSFGs provides a promising way of studying the $z$$>$3 tail of DSFGs, including their most distant ($z$$>$5) counterparts.

This sample of SPT-DSFGs is the most complete spectroscopic sample of DSFGs in the literature. Besides studying the redshift distribution of DSFGs, spectroscopic redshifts are an important first step for future detailed studies of the ISM at high redshifts \citep[e.g.,][]{aravena13, bothwell13, gullberg15, spilker15}.
In the future, we will work towards our goal of obtaining redshifts for the complete sample of 100 SPT-DSFGs, which will enable detailed studies of the ISM over cosmic time. 
\\

\acknowledgments 											 %
MLS was supported for this research through a stipend from the International Max Planck Research School (IMPRS) for Astronomy and Astrophysics at the Universities of Bonn and Cologne. M.A. acknowledges partial support from FONDECYT through grant 1140099. JDV, KCL, DPM, and JSS acknowledge support from the U.S. National Science Foundation under grant No. AST-1312950.
This paper makes use of the following ALMA data:
ADS/JAO.ALMA\# 2012.1.00844.S, 2012.1.00994.S, 2011.0.00957.S and 2011.0.00958.S. ALMA is a partnership of ESO (representing its member states), NSF (USA) and NINS (Japan), together with NRC (Canada) and NSC and ASIAA (Taiwan), in cooperation with the Republic of Chile. The Joint ALMA Observatory is operated by ESO, AUI/NRAO and NAOJ.
This work is based in part on observations made with {\it Herschel} under program ID's OT1\_jvieira\_4 and DDT\_mstrande\_1. {\it Herschel} is a European Space Agency Cornerstone Mission with significant participation by NASA. We also use data from the Atacama Pathfinder Experiment under program IDs E-086.A-0793A-2010, M-085.F-0008-2010, M-087.F-0015-2011, M-091.F-0031-2013, E-094.A-0712A-2014, M-095.F-0028-2015, E-096.A-0939A-2015. APEX is a collaboration between the Max-Planck-Institut f\"{u}r Radioastronomie, the European Southern Observatory, and the Onsala Space Observatory.
The Australia Telescope is funded by the Commonwealth of Australia for operation as a National Facility managed by CSIRO. We have also used data from VLT/X-Shooter under the ESO project ID E-092.A-0503(A).
The SPT is supported by the National Science Foundation through grant PLR-1248097, with partial support through PHY-1125897, the Kavli Foundation and the Gordon and Betty Moore Foundation grant GBMF 947.

\bibliographystyle{apj}										 %
\bibliography{spt_smg}										 %

\appendix													 %

\section{Supplementary redshift information on sources from Wei$\ss$ \etal\  2013}
\label{sect:supplementary_redshift_info}

We show here the supplementary observations that resolve redshift ambiguities in the ALMA observations from \citet{weiss13} and go through the lines found in the ALMA 1\,mm observations.
\\

\emph{\textbf{SPT0125-50:}}
The most likely redshift option from \citet{weiss13} was confirmed by an H$_2$O absorption line from our ALMA 870\,\micron\ high resolution imaging cube for this source, presented by \citet{spilker14}, along with a CO(10--9) detection at 232.35\,GHz and H$_2$O emission lines from the ALMA 1\,mm data. These detections identify the original ALMA lines as CO(4--3) and \ci\ at redshift $z$=3.959(3). 
We only clearly see one H$_2$O line, as one line blends with the CO line and one is at the edge of the spectrum.
\\

\emph{\textbf{SPT0300-46:}}
In \citet{weiss13}, one line was detected and one tentative line feature was seen. The most likely line identification was found to be CO(4--3) and \ci\ with a redshift of $z$=3.5954(7). This was confirmed by observations of \cii\ with APEX, presented in \citet{gullberg15}, along with the CO(10--9) line at 250.71\,GHz and the H$_2$O$^+$($1_{11}-1_{00}$, $J_{3=1/2-1/2}$) absorption line at 247.97\,GHz in the ALMA 1\,mm spectrum. 
\\

\begin{figure}[t]
	\centering
	\includegraphics[viewport=0 0 330 335, clip=true,height=5cm,angle=0]{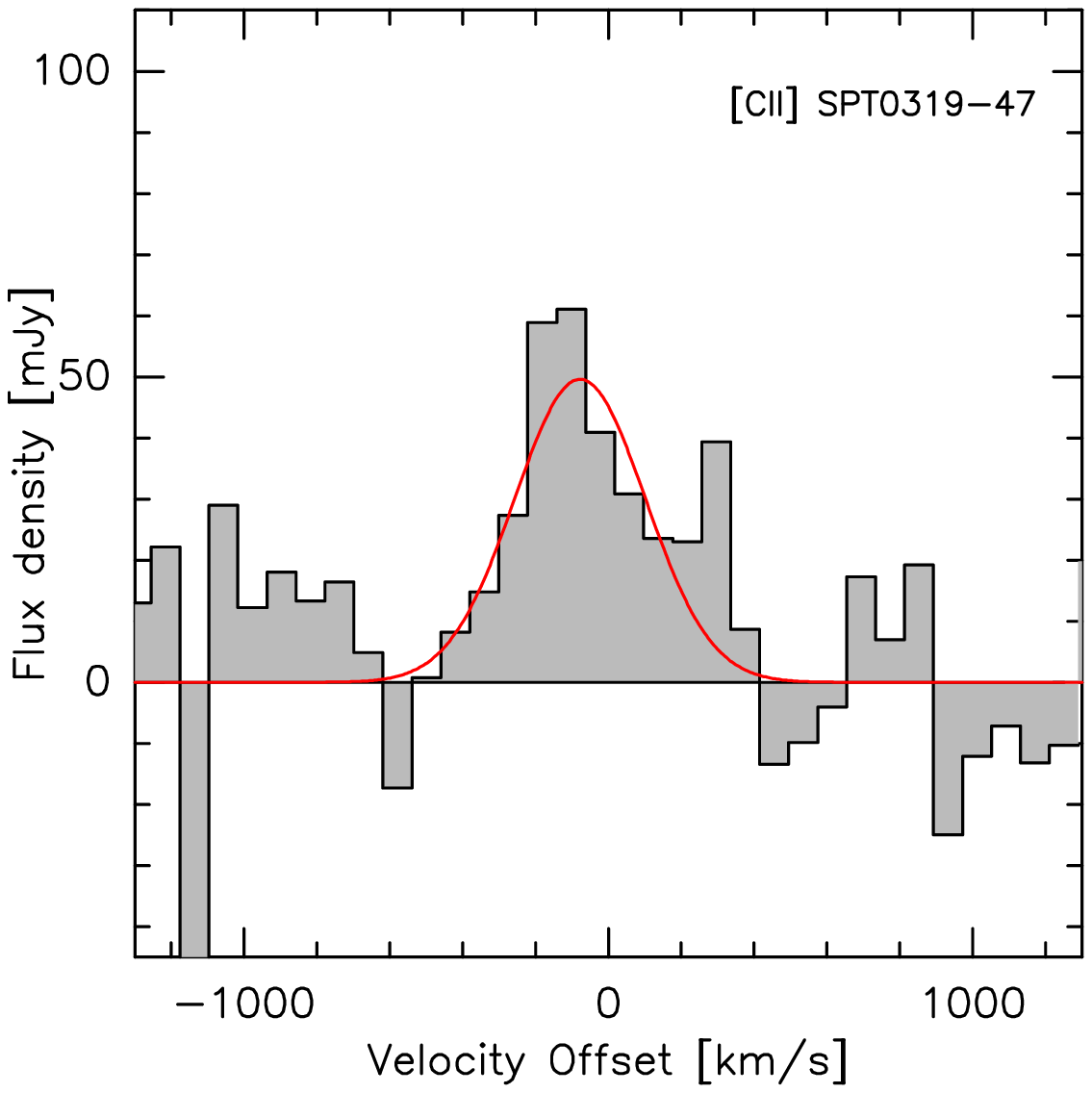} \hspace{0.5cm}
	\includegraphics[viewport=0 0 330 335, clip=true,height=5cm,angle=0]{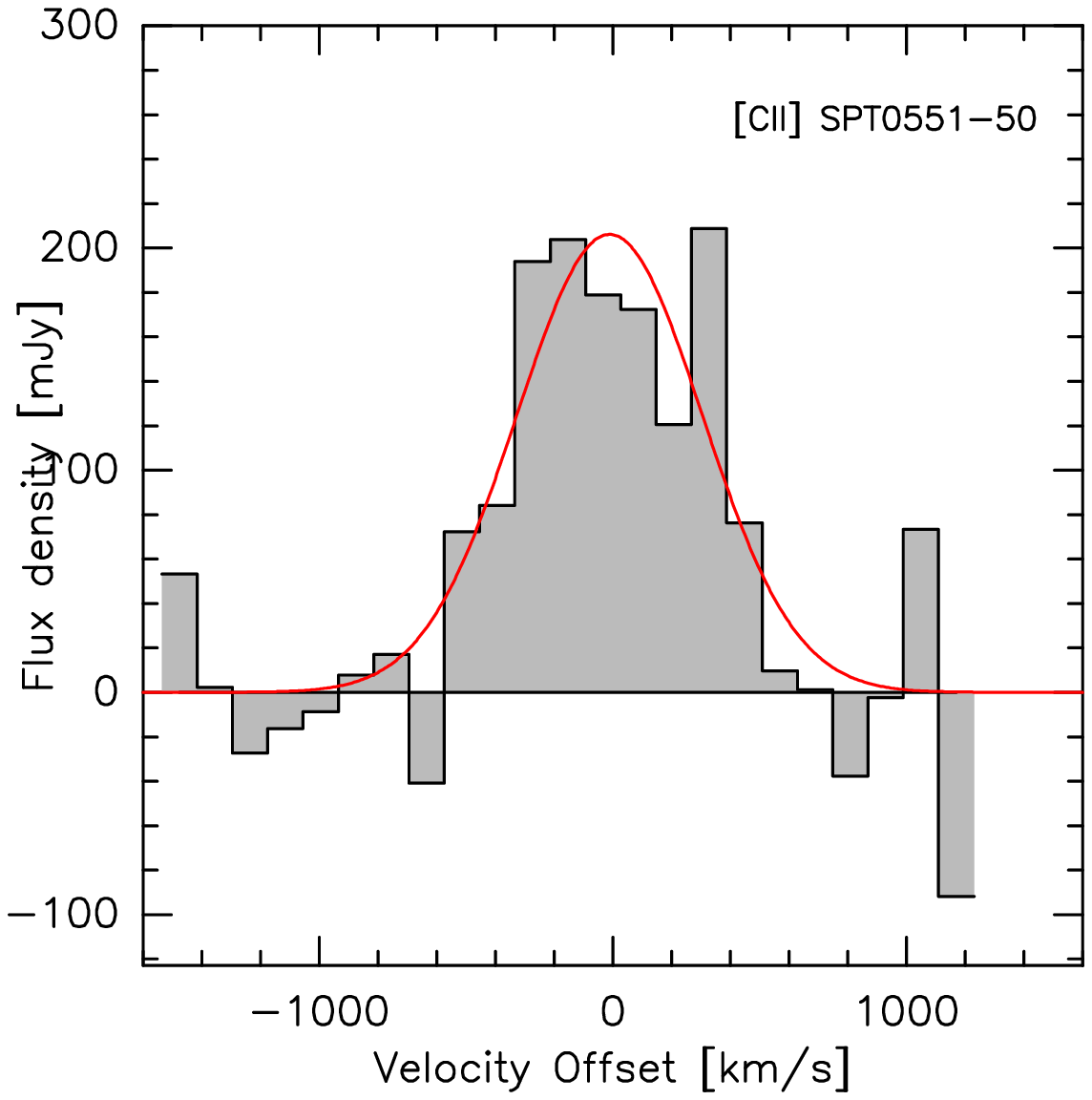} %
	\caption{APEX/FLASH observations of \cii . The observations are described in Section \ref{sect:add-spec}. \emph{Left}: SPT0319-47, for which \cii\ cements the redshift. \emph{Right}: SPT0551-50, where the redshift was wrongly determined using optical spectroscopy but for which this \cii\ line now robustly determines the redshift.}
	\label{Fig:CII-lines}
\end{figure}

\emph{\textbf{SPT0319-47:}} 
\citet{weiss13} present this source as having no lines though the ALMA 3\,mm spectrum shows a wide (FWHM$\sim$1700\,km\,s$^{-1}$) tentative line at 104.39\,GHz, with the most probable identification being CO(5--4) at $z$=4.516(4).
The redshift was confirmed by the CO(12--11) line at 250.77\,GHz in the 1\,mm ALMA spectrum. In this source we have also detected \cii\ with APEX/FLASH, see Figure \ref{Fig:CII-lines}.
\\

\emph{\textbf{SPT0441-46:}} 
The redshift of this source was confirmed with APEX/FLASH \cii\ observations before the publication of \citet{weiss13}, but it was not confirmed by the submission of the targeted 1\,mm redshift confirmation proposal. This source had two likely redshift options and it was therefore observed in two tunings. 
In the 1\,mm data we see a double peaked CO(11--10) line at 231.19\,GHz and a double peaked H$_2$O($2_{20}-2_{11}$) line at 224.33\,GHz. We also detect NH$_3$, in the form of NH$_3$($2_0 - 1_0$). 
\\

\begin{figure}[t]
	\centering
	\includegraphics[viewport=0 0 410 215, clip=true,width=7cm,angle=0]{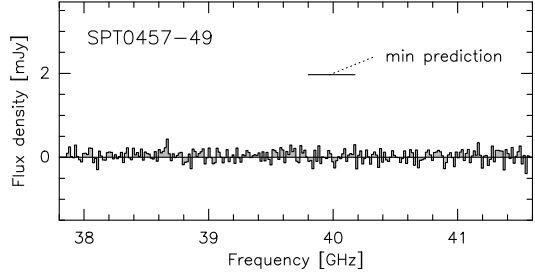} 
	\caption{An ATCA spectrum of SPT0457-49 looking for CO(1--0) at $1.77$$<$$z$$<$$2.0$ in the redshift desert of the 3\,mm ALMA spectral scans. }
	\label{Fig:SPT0457-zdesert}
\end{figure}

\emph{\textbf{SPT0457-49:}}
\citet{weiss13} did not find any lines in this source and assumed it was in the redshift desert. With ATCA we have scanned the redshift range 1.77$<$$z$$<$2.05 searching for CO(1--0) without success (see Figure \ref{Fig:SPT0457-zdesert}). We are no closer to determining the spectroscopic redshift of this source, though it is clear that its redshift cannot be assumed to lie in the redshift desert as suggested by \citet{weiss13} and with $z_{phot}$=3.4$\pm$0.6 it probably does not.
\\

\emph{\textbf{SPT0459-58:}}
\citet{weiss13} present a single line with two almost equally likely redshift options for this source. The highest redshift option at $z$=4.856(4) with the line identification CO(5--4) was confirmed by the CO(11--10) line at 216.36\,GHz in the 1\,mm ALMA spectrum.
\\

\begin{figure*}[htb]
	\centering
	\includegraphics[viewport=0 50 360 380, clip=true,width=5cm,angle=0]{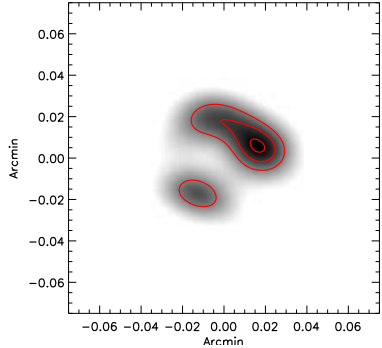}
	\caption{The ALMA 1\,mm continuum imaging of SPT0512-59. The \emph{red} contours show 5,7 and 9\,$\sigma$ and indicate where the brightest component is found. }
	\label{Fig:SPT0512-highres}
\end{figure*}

\emph{\textbf{SPT0512-59:}}
One line  with two possible identifications is presented by \citet{weiss13}.
The most likely of these, CO(3--2) at $z$=2.2331(2), was confirmed by the detection of \cii\ with SPIRE FTS, presented by \citet{gullberg15}.
This source was also observed with ALMA at 1\,mm where we detected the CO(6--5) line at 213.89\,GHz and at this high resolution the lens is resolved, see Figure \ref{Fig:SPT0512-highres}. The \emph{red} contours mark 5,7 and 9\,$\sigma$. We extracted the spectrum where we found the highest SNR.
\\

\emph{\textbf{SPT0550-53:}}
A single line with two possible identifications was presented for this source in \citet{weiss13}, where the most likely line identification, CO(4--3) at $z$=3.1280(7) was confirmed by a \cii\ detection from APEX \citep[see][]{gullberg15}, along with a double peaked CO(8--7) line at 223.31\,GHz and a double peaked H$_2$O($2_{02}-1_{11}$) line at 239.36\,GHz in the 1\,mm ALMA spectrum. 
\\

\begin{figure}
	\centering
	\includegraphics[viewport=0 0 340 330, clip=true,height=5cm,angle=0]{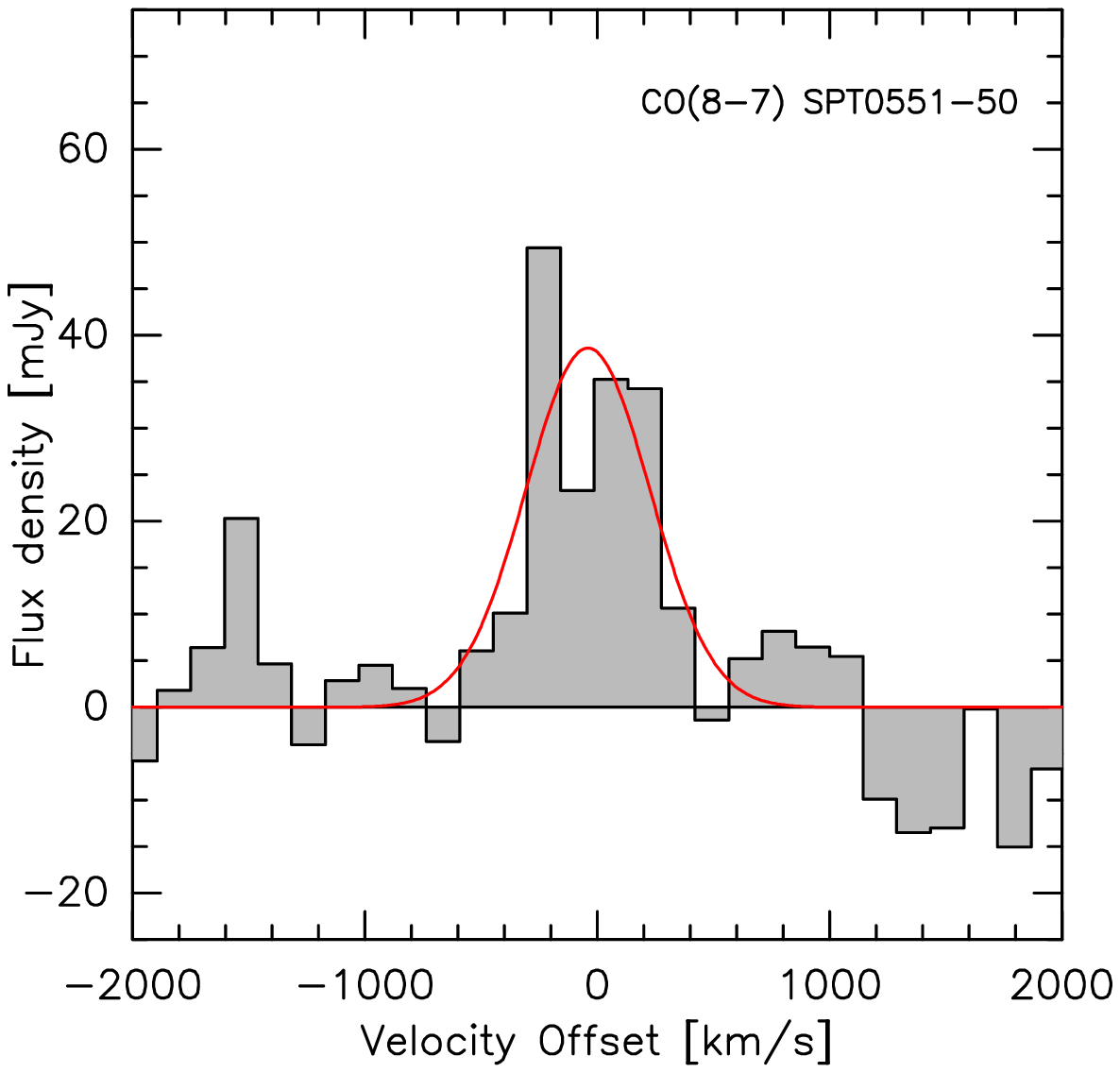} \hspace{1cm}
	\includegraphics[viewport=0 0 340 330, clip=true,height=5cm,angle=0]{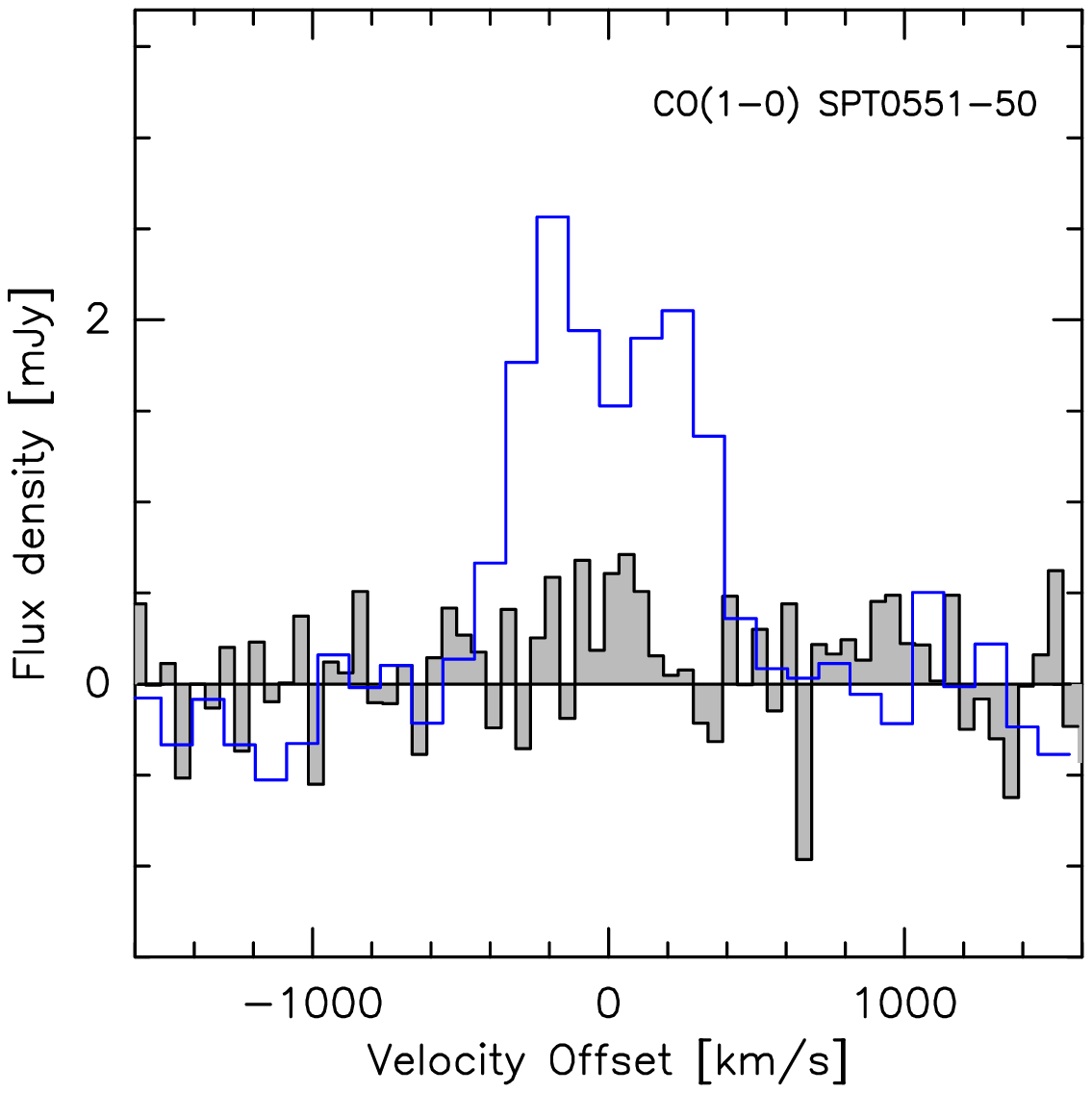} 
	\caption{The lines helping in the redshift identification of SPT0551-50. \emph{Left}:  APEX/SHeFI observation of the CO(8--7) line (See Gullberg \etal\ in prep.). \emph{Right}: ATCA CO(1--0) non detection. The \emph{blue} line shows the line from the ALMA 3\,mm spectrum scaled to the expected CO(1--0) flux density.}
	\label{Fig:SPT0551-50}
\end{figure}

\emph{\textbf{SPT0551-50:}}
This source was presented by \citet{weiss13} as a having a secure redshift of $z$=2.1232(2). This was based on a combination of a line in the ALMA 3\,mm spectrum identified as CO(3--2) and a VLT C{\scriptsize IV} line. The source has since been followed up on with several facilities: \citet{gullberg15} present a \cii\ non-detection observed with {\it Herschel}/SPIRE FTS and CO(1--0) observations with ATCA also showed a non-detection (see the right panel of Figure \ref{Fig:SPT0551-50}). APEX/SHeFi observations of the CO(6--5) yielded a detection though (see left panel of Figure \ref{Fig:SPT0551-50}), these data will be presented by Gullberg \etal\ in prep.
We investigated the possibility of the VLT line belonging to a foreground system and left it out of the redshift predictions following here. This opens up the option for the ALMA 3\,mm CO line to be identified as CO(4--3) and the APEX/SHeFi CO line to be CO(8--7) at $z$=3.1638(3). The shift in frequency would be so small between the previous CO(6--5) identification and the new CO(8--7) identification that we would not be able to detect the difference with the spectral resolution of APEX/SHeFI.
The photometry strongly favors the redshift $z$=3.1638(3) for which we find the dust temperature  $T_{\rm dust}$=37$\pm$1\,K whereas the lower redshift option yields a dust temperature of $T_{\rm dust}$=27$\pm$1\,K. The photometric redshift for this source is $z_{phot}$=3.1$\pm$0.6.
This redshift option was confirmed by the detection of \cii\ with APEX/FLASH, see Figure \ref{Fig:CII-lines}.
\\

\emph{\textbf{SPT2132-58:}} 
The redshift for this source was already confirmed by the time of publication of \citet{weiss13} through \cii\ observations with APEX but it had already been included in the ALMA 1\,mm follow up project, where we then detected CO(12--11) at 239.59\,GHz and \nii\ at 253.32\,GHz.
\\

\section{Supplementary information for new sources presented in this work}
\label{sect:supplementary_singleline}

Below we discuss the eight individual cases which have zero or one CO line detected in the new ALMA 3\,mm data along with SPT2357-51 for which we have additional optical observations. We also show the \cii\ spectra obtained with APEX/FLASH and present an APEX/Z-Spec spectrum that was used to find the redshift of SPT0551-48 for which we do not have ALMA observations.\\

\begin{figure*}[htb]
	\centering
	\includegraphics[width=17cm,angle=0]{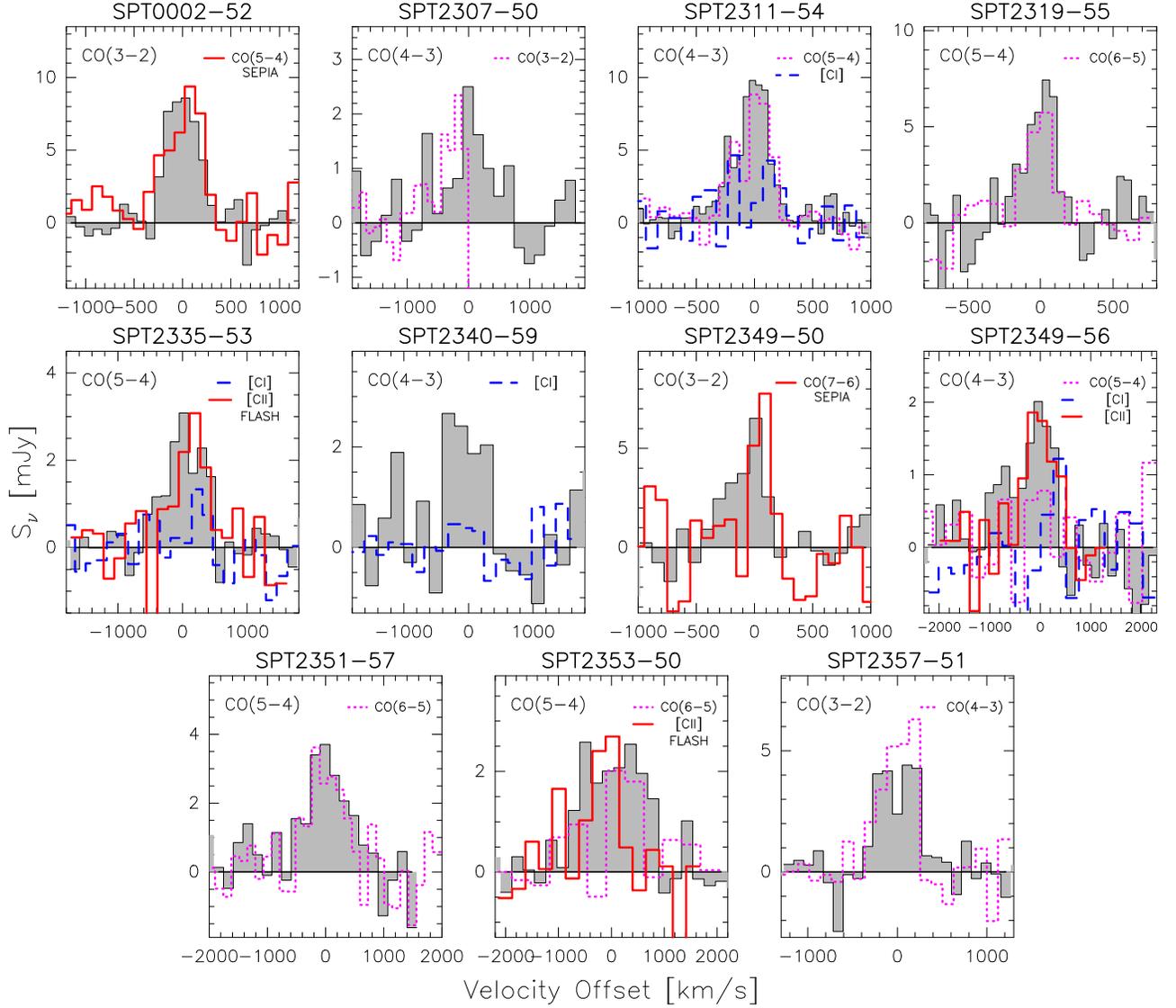}
	\caption{Line overlays of all sources showing one or more line in the ALMA 3\,mm spectra.
	In the top left corner of each plot the line shown with a \emph{grey} histogram is given. Other CO lines are overlaid in \emph{pink}, \ci\ lines are overlaid in \emph{blue} and lines obtained with APEX are overlaid in \emph{red} and the instrument is given with the line name.
	All ALMA lines are shown at their true flux density, but the lines observed with APEX/SEPIA and APEX/FLASH have been scaled so they could be shown in the same plot. Note that both the velocity axis and the flux density axis varies over the sources as they have been adjusted to better show the lines.
	For SPT2307-50, the CO(3-2) line is at the edge of the spectrum, which is why it stops mid-line. Four sources (SPT2307-50, SPT2340-59, SPT2349-50, SPT2349-56) still have ambiguous redshifts and are here shown at their most probable redshift.
	For details on the sources, see the source description in this Appendix. }
	\label{Fig:line-overlay}
\end{figure*}

\emph{\textbf{SPT0002-52:}}
We detect a single line at 103.19\,GHz, which turned out to be CO(3--2) at  $z$=2.3510(2) ($T_{\rm dust}$=42$\pm$2\,K).
This was confirmed with APEX/SEPIA where we detected the CO(5--4) line at 171.97\,GHz, see Figure \ref{Fig:line-overlay}.
\\

\begin{figure*}[htb]
	\centering
	\includegraphics[width=8cm,angle=0]{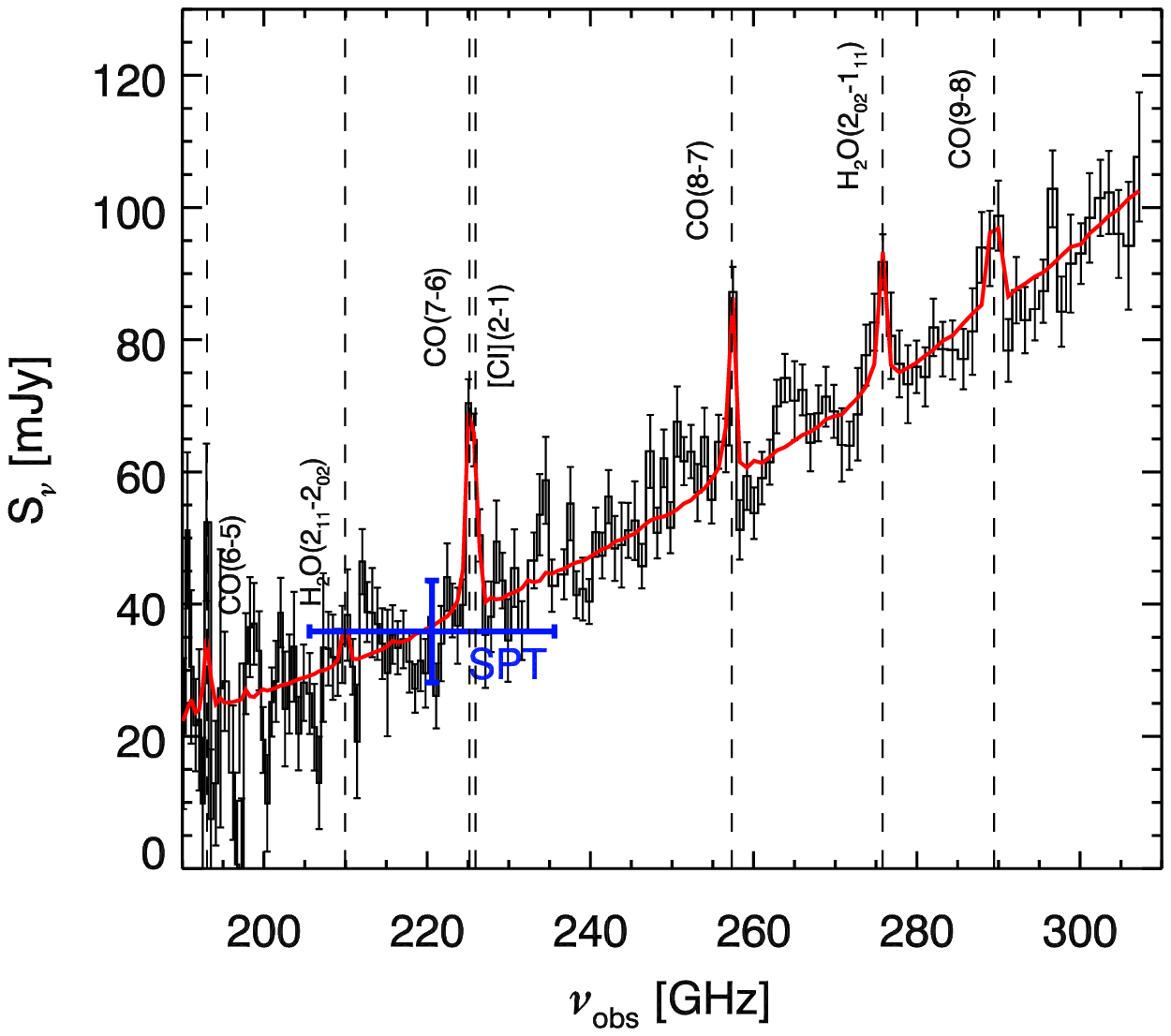}
	\caption{APEX/Z-Spec spectrum of SPT~0551-48. Four lines, CO(7--6), CO(8--7), \ci (2--1) and H$_2$O($2_{01} - 1_{11}$), secure the redshift to $z$=2.579.}
	\label{Fig:SPT0551Zspec}
\end{figure*}

\emph{\textbf{SPT0551-48:}}
This source was not in the ALMA redshift search. Instead a redshift search was performed with APEX/Z-Spec, see Figure \ref{Fig:SPT0551Zspec} for the spectrum and Section \ref{Sect:Zspec-obs} for a description of the data. In the spectrum we find at least four lines, CO(7--6), CO(8--7), \ci (2--1) and H$_2$O. Furthermore the CO(1--0) line was detected for this source using ATCA and improving the precision on the redshift, finding $z$=2.5833(2) (\citet{aravena15}.).
\\

\emph{\textbf{SPT2307-50:}}
In this source we find a weak line at 112.30\,GHz. We exclude the line identification CO(5--4) at $z$=4.132(4), as we would see CO(4--3) in the observing window. If the line is CO(4--3) at $z$=3.105(2), the CO(3--2) line would fall just below the frequency range of the observing window. When the spectrum is smoothed as in Figure \ref{Fig:line-overlay} we do not see anything, but when investigating the edge of the spectrum unsmoothened, we find indication for the rise of a line. Figure \ref{Fig:line-overlay} shows the possible side of the CO(3--2) line overlaid on the CO(4--3) line. Since this is not a clear detection we still consider the line identification CO(3--2) at $z$=2.079(1) which would have $T_{\rm dust}$=25$\pm$3\,K. CO(2--1) at $z$=1.052(1) is ruled out since the dust temperature ($T_{\rm dust}$=16$\pm$1\,K) would be too low. The most probable line identification based on the photometric redshift $z_{phot}$=3.4$\pm0.9$ is CO(4--3) which would then have a dust temperature of $T_{\rm dust}$=36$\pm$4\,K.
\\

\emph{\textbf{SPT2335-53:}}
We detect a line at 100.12\,GHz and a tentative feature at 85.51\,GHz, which turns out to be CO(5--4) and \ci\ at $z$=4.755(1). 
This was confirmed by a \cii\ detection from APEX/FLASH, see Figure \ref{Fig:line-overlay}. At this redshift we find a dust temperature of $T_{\rm dust}$=57$\pm$4\,K.
\\

\begin{figure*}[htb]
	\centering
	\includegraphics[viewport=0 0 445 410, clip=true,width=6cm,angle=0]{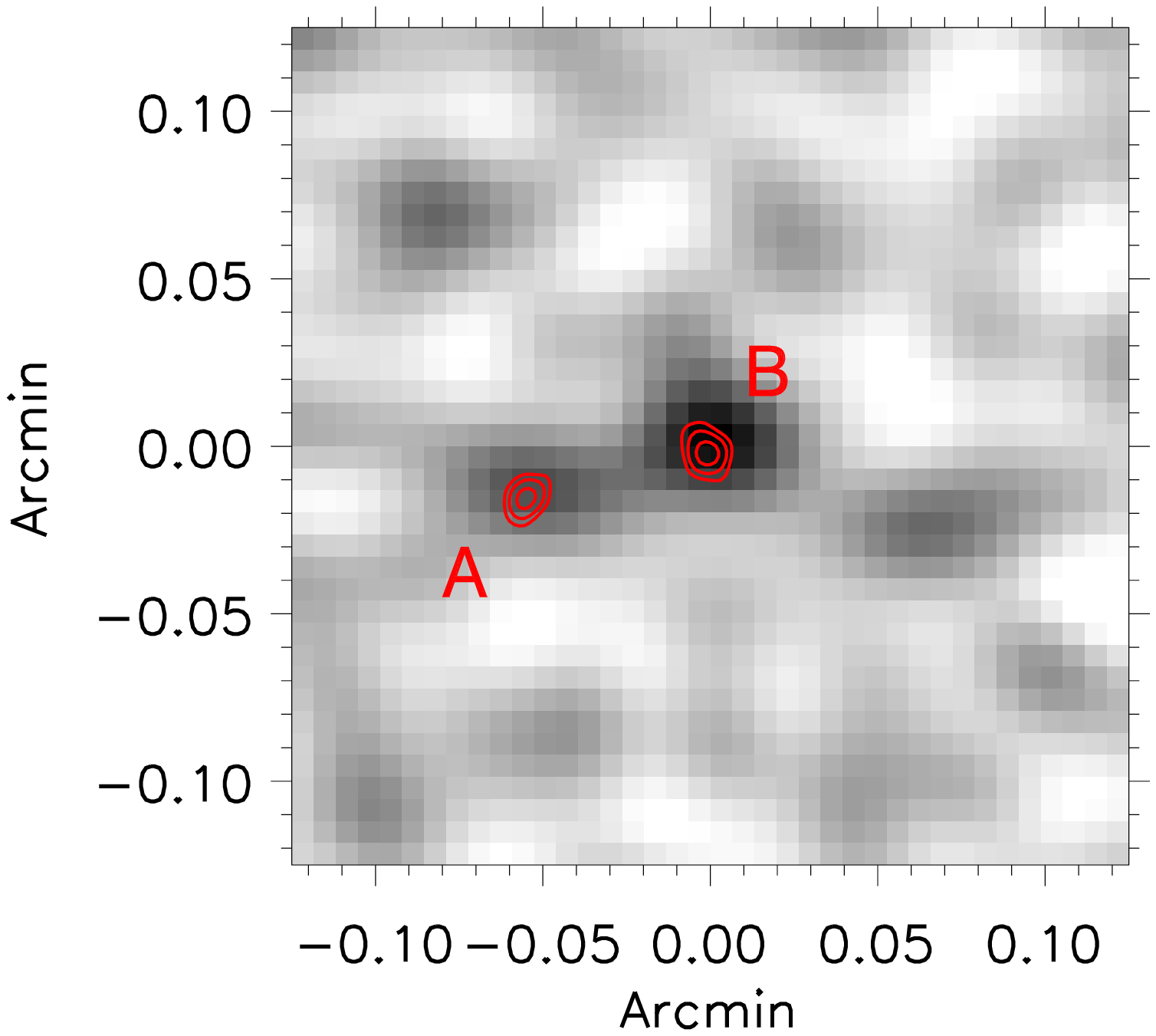} \hspace{1.5cm}
	\includegraphics[viewport=0 0 445 410, clip=true,width=6cm,angle=0]{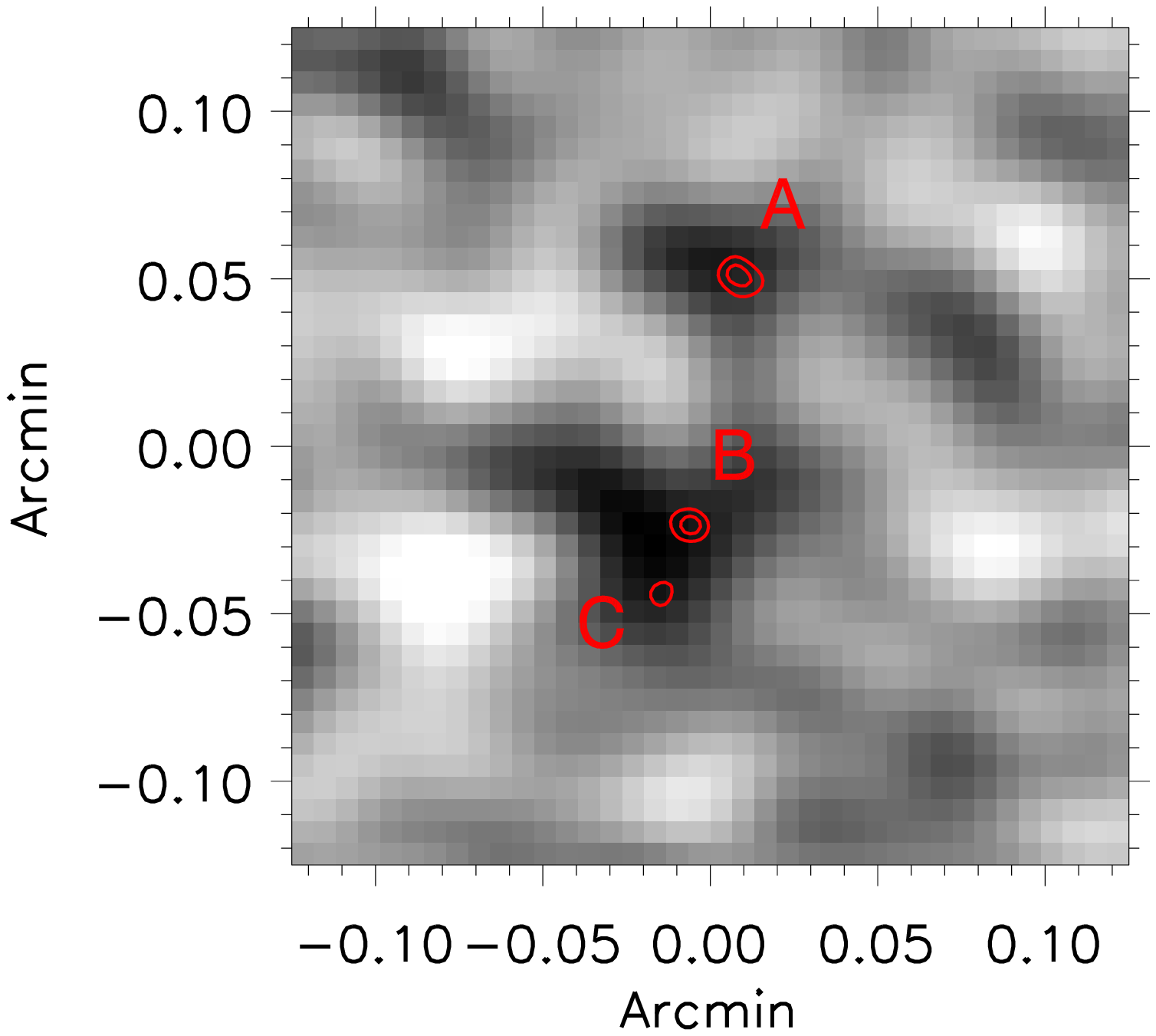}
	\caption{3\,mm continuum imaging of the two sources that split up in to multiple components scaled by the rms (which is rms$=$0.05mJy for both sources). The contours are high resolution 870\,$\mu$m ALMA high resolution imaging, 5,10,20,30$\sigma$. \emph{Left}: SPT2340-59 which splits up in to two counterparts, with the peak flux density of $S_{peak}$=0.30\,mJy. \emph{Right}: SPT2349-56  splits into three at 870\,$\mu$m, with the peak flux density of $S_{peak}$=0.22\,mJy.}
	\label{Fig:MulComIm}
\end{figure*}

\emph{\textbf{SPT2340-59:}}
This source splits up into two counterparts in the ALMA 3\,mm continuum image. In our 870\,$\mu$m high resolution ALMA imaging, we see the same two counterparts (see Figure \ref{Fig:MulComIm}). Counterpart B is brightest, but in the spectrum of this we do not see any lines. In counterpart A, however, we see a potential line at 94.79\,GHz. If this line is real the possible line identifications are CO(3--2) at $z$=2.6480(8), CO(4--3) at $z$=3.864(1) or CO(5--4) at $z$=5.079(1). CO(2--1) at $z$=1.4321(5) is excluded as that would mean a dust temperature of $T_{\rm dust}$=17$\pm$1\,K, which has not been observed in any of our sources. 
The photometric redshift of the source is $z_{phot}$=3.8$\pm$0.7 favoring the CO(4--3) line identification. For this redshift though, the \ci\ line falls within the spectral window. With the low SNR of the CO line, it is reasonable to assume that the \ci\ line is hiding within the noise.
\\

\emph{\textbf{SPT2349-50:}}
In this source we see a single bright line at 89.21\,GHz with the most probable line identification being CO(3--2) at $z$=2.8764(3). This was confirmed by APEX/SEPIA observations of the CO(7--6) line at 207.99\,GHz. 
\\

\begin{figure*}[htb]
	\centering
	\includegraphics[viewport=0 0 515 205, clip=true,width=12cm,angle=0]{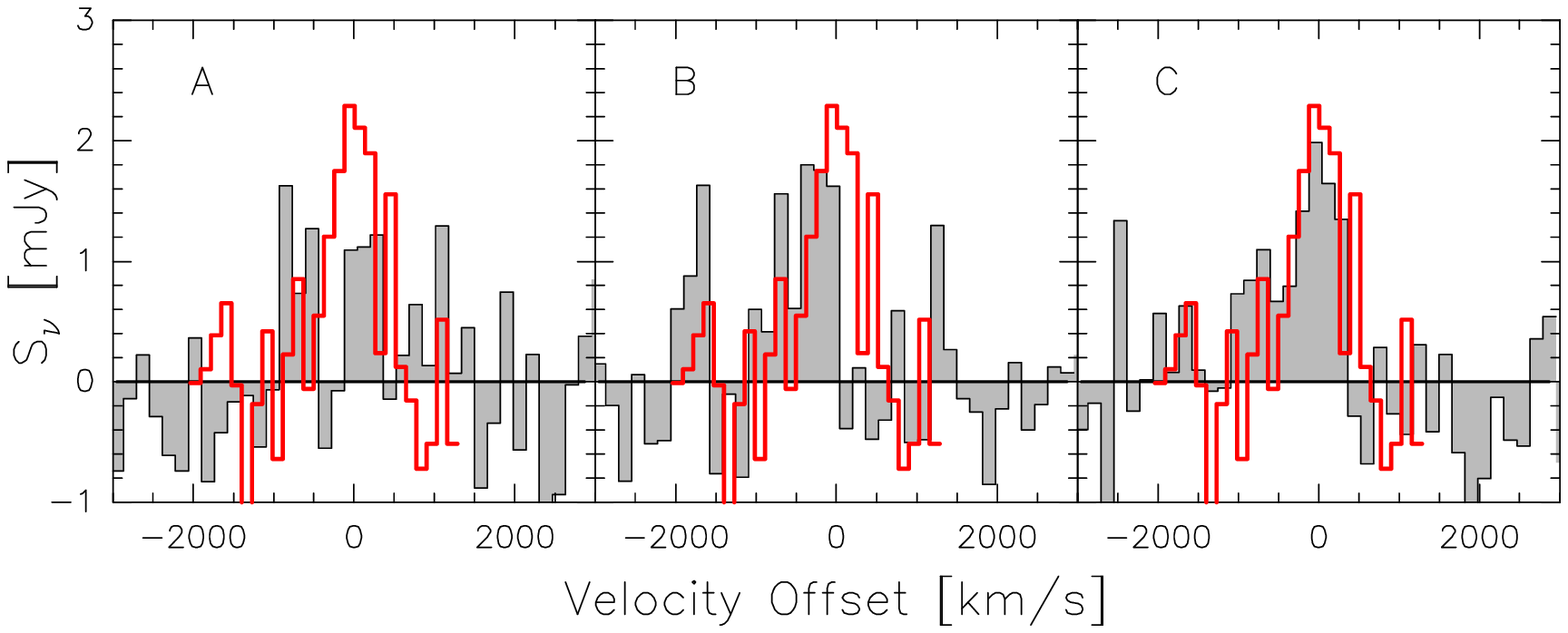}
	\caption{The three panels shows the CO(4--3) extracted at position A, B and C, with the \cii\ overlaid in red.}
	\label{Fig:SPT2349-56-panels}
\end{figure*}

\emph{\textbf{SPT2349-56:}}
At 3\,mm, this source splits into two counterparts, but in the 870\,$\mu$m high resolution ALMA imaging \citep{vieira13} we see three counterparts (see Figure \ref{Fig:MulComIm}). We use the high resolution imaging to define the positions for the three counterparts and extract spectra there. We see indications of a line at $\sim$87.0GHz all positions (with the center slightly shifted at each position), with position A showing the weakest line. 
When the spectra of these three positions are stacked we see a hint of a line at 108.62GHz.
These two lines are consistent with the line identifications CO(4--3) and CO(5--4) at z=4.306(?) which is confirmed by APEX/FLASH \cii\ observations. The \cii\ only traces part of the line seen in the stack (see Figure \ref{Fig:line-overlay}), so in Figure \ref{Fig:SPT2349-56-panels} we show the spectra of each of the components with \cii\ overlaid. It is clear from this that the \cii\ traces component C.
\\

\emph{\textbf{SPT2353-50:}}
We detect a single wide spectral feature at  87.63\,GHz. We rule out the line identification CO(2--1) at $z$=1.630(2) and CO(3--2) at $z$=2.945(2) because the dust temperatures would be $T_{\rm dust}$=16$\pm$1\,K and $T_{\rm dust}$=25$\pm$2\,K respectively. 
The photometric redshift $z_{phot}$=4.5$\pm$0.8 favors the line identification CO(4--3) at $z$=4.261(3). 
The last possible line identification is CO(5--4) at $z$=5.576(4) which is not negligible with a dust temperature of $T_{\rm dust}$=46$\pm$2\,K. At this redshift, CO(6--5) falls within the spectral window, and we see a SNR$\sim$1.5 feature at the frequency where the line should fall. This redshift option was confirmed by the detection of \cii , see Figure \ref{Fig:line-overlay}.
\\

\begin{figure*}[htb]
	\centering
	\includegraphics[viewport=0 0 330 330, clip=true,height=5cm,angle=0]{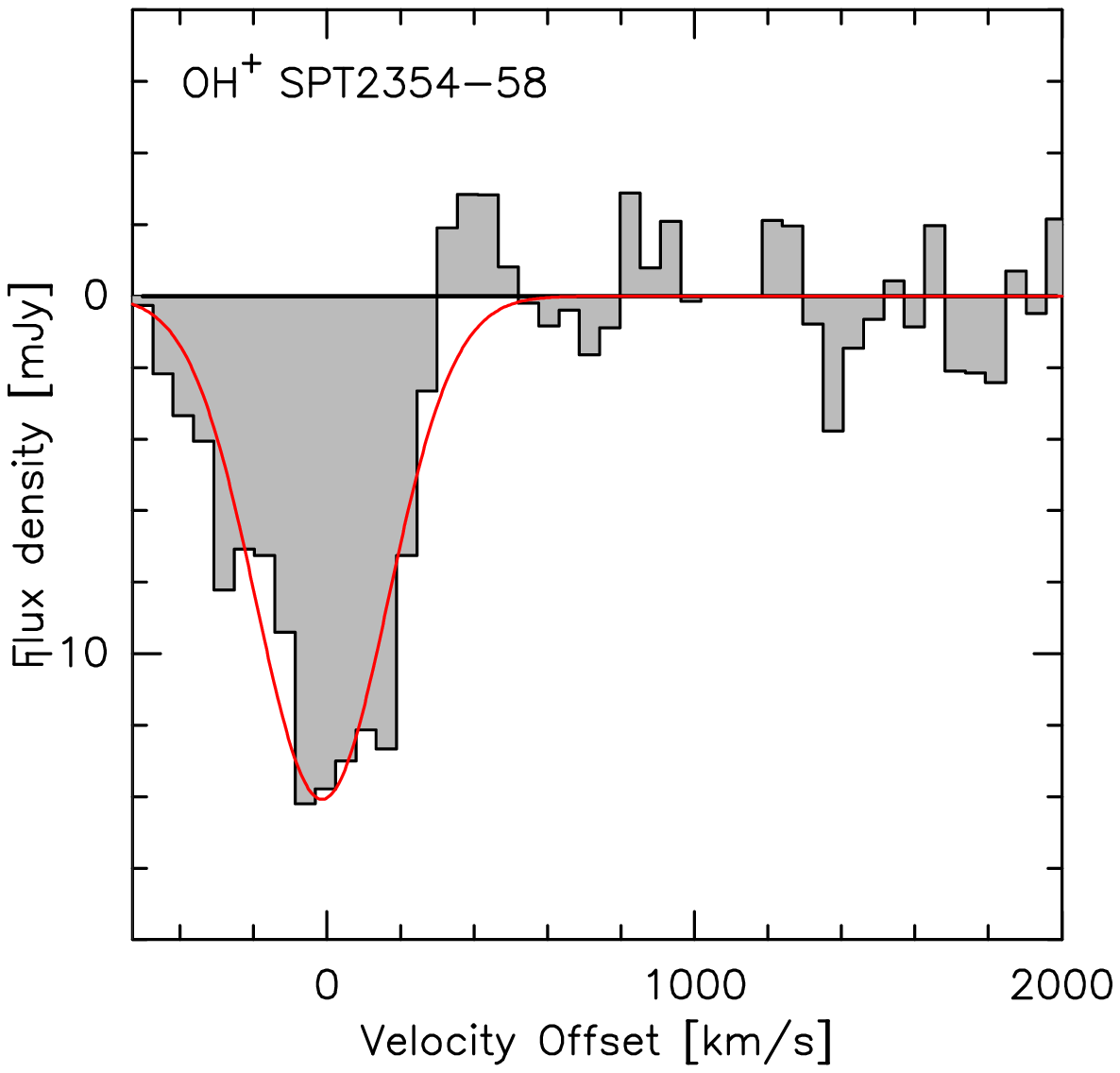} \hspace{1cm}
	\includegraphics[viewport=0 0 330 330, clip=true,height=5cm,angle=0]{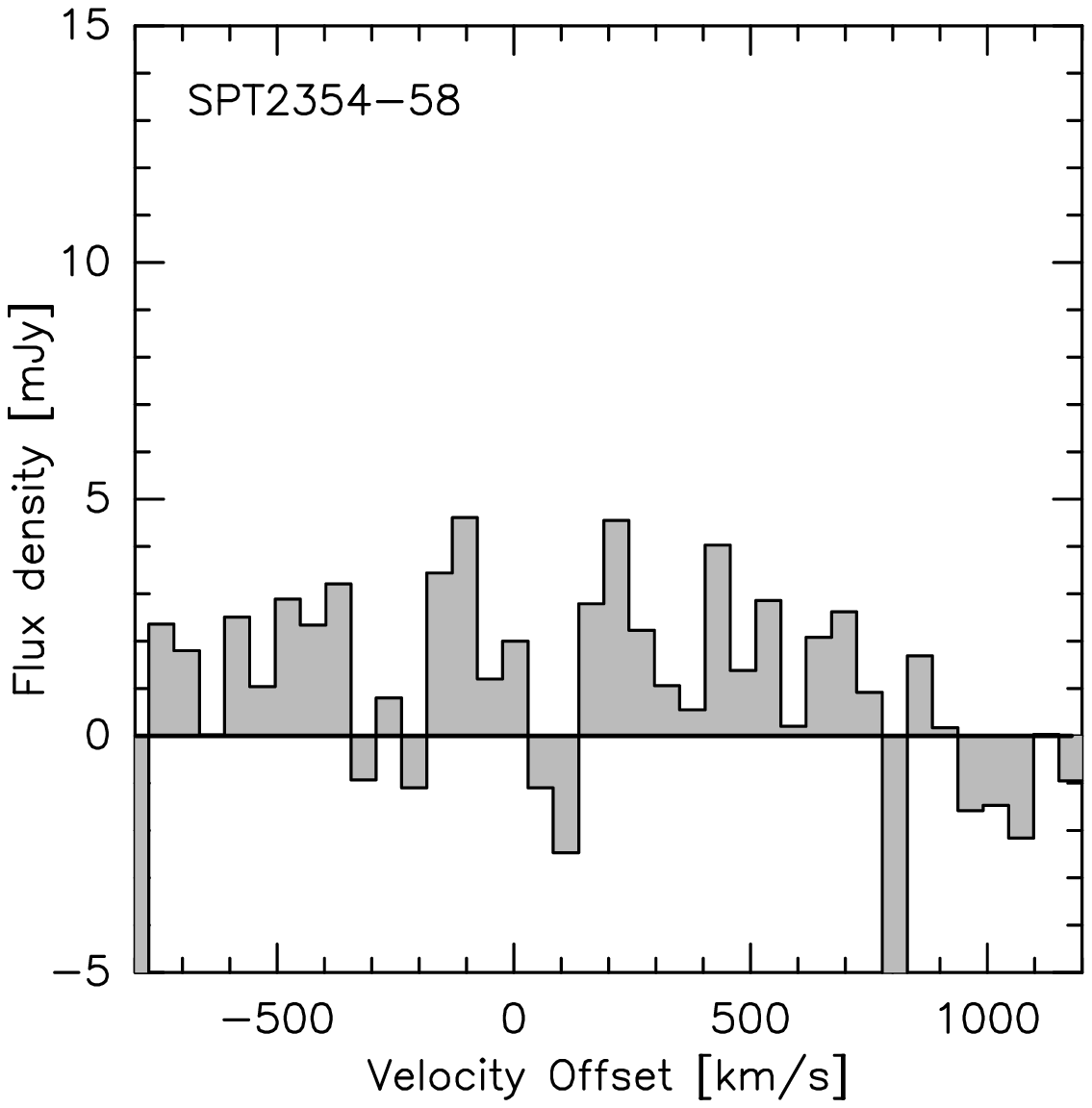}
	\caption{These two spectra are extracted from the high resolution 870\,\micron\ data cube available for SPT2354-58. \emph{Left}: Absorption line with OH$^+$ as the most likely line identification.  \emph{Right:} Where CO(5--4) would have been if the redshift is $z$=0.6431(3).}
	\label{Fig:SPT2354-58}
\end{figure*}

\emph{\textbf{SPT2354-58:}}
This is the only source where we do not find any lines in the 3\,mm ALMA redshift search. For this source we have high resolution 870\,\micron\ imaging and in this data cube we found an absorption line at 338.95\,GHz (see left panel of Figure \ref{Fig:SPT2354-58}). We identify the line as either OH$^+$(1$_{22}$ -- 0$_{11}$) at $z$=1.867(1) or H$_2$O($1_{10}-1_{01}$) at $z$=0.6431(3). For other absorptions line identifications we should have seen an emission line in the 3\,mm ALMA data. The first option is favored by the photometry with $T_{\rm dust}$=43$\pm$2\,K compared to $T_{\rm dust}$=27$\pm$1\,K for the second option. Furthermore we should have seen CO(5--4) at 350.77\,GHz if the second option was correct and this is not the case (see right panel of Figure \ref{Fig:SPT2354-58}). The most probable redshift is therefore identified as $z$=1.867(1).
\\

\begin{figure*}[htb]
	\centering
	\includegraphics[viewport=0 0 580 280, clip=true,height=4.5cm,angle=0]{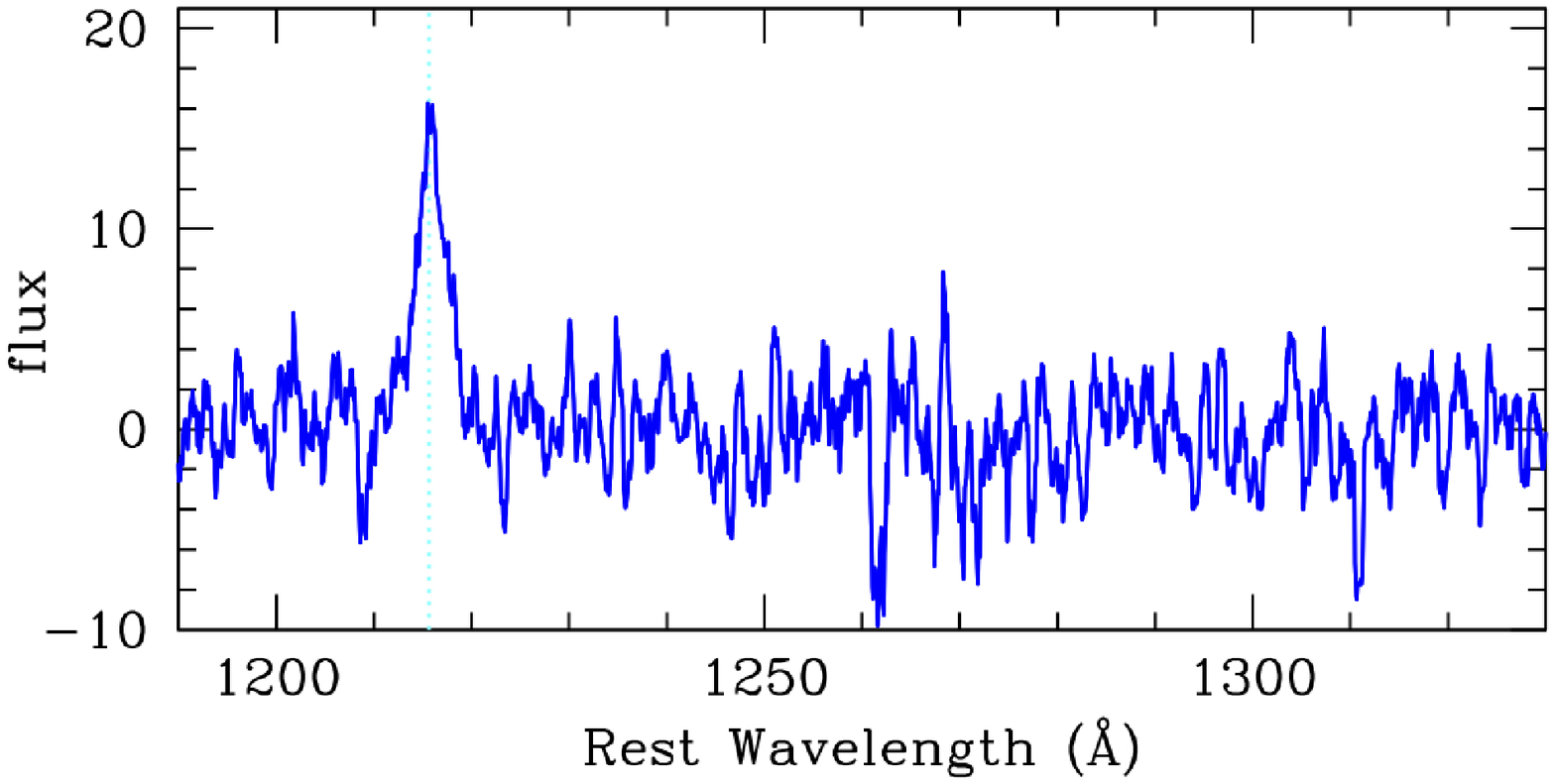} 
	\caption{UV/Optical/near-IR spectra (VLT/X-shooter) of the $z$=3.07 identified SPT2357-51. ÊThe redshift was identified with large equivalent width Lyman-$\alpha$ emission in the UV, exhibiting a strong, broad emission line (FWHM = 1230 km/s) but no detectable continuum. 
	The broad Ly$\alpha$ line is suggestive of an AGN, although no CIV$_{1549 \AA}$ nor OIII$_{5007 \AA}$ are detected.}
	\label{Fig:SPT2357-51}
\end{figure*}

\emph{\textbf{SPT2357-51:}}
For this source optical spectroscopy with the X-shooter/VLT was performed before the ALMA observations, detecting a line. The observations are described in Section \ref{Sect:Anc_spec} and the spectrum is shown in Figure \ref{Fig:SPT2357-51}.

\section{Supplementary Far-Infrared Photometry}
\label{sect:supplementary_photometry}
 In this Appendix we show Table \ref{Tab:photometry} which contains the values obtained from the photometric observations described in Section \ref{Sect:Photometry}.
 
\begin{deluxetable}{c c c c c c c c c c c}
\tabletypesize{\footnotesize}
\tablecaption{Photometry of all sources\label{Tab:photometry}}
\tablewidth{0pt}
\tablehead{
\colhead{Source}      & \colhead{S3000/mJy} & \colhead{S2000/mJy} & \colhead{S1400/mJy} & 
\colhead{S870/mJy} & \colhead{S500/mJy} & \colhead{S350/mJy}   & \colhead{S250/mJy}    &
\colhead{S160/mJy} & \colhead{S100/mJy} 
}
\startdata
SPT0002-52 & 	0.44$\pm$0.05	& 2.9$\pm$1.3 & 12.5$\pm$5.4 	& 50.3$\pm$3.8 	& 202.0$\pm$10.0 	& 283.5$\pm$8.9 	& 332.9$\pm$10.1 	& 234$\pm$21 		& 94$\pm$5 		\\
SPT2307-50 & 	0.26$\pm$0.05	& 1.2$\pm$1.4 & 5.8$\pm$6.7 		& 22.1$\pm$2.8 	& 37.6$\pm$10.4 	& 42.3$\pm$11.2 	& 50.5$\pm$12.2 	&   				&   				\\
SPT2311-54 & 	0.55$\pm$0.05	& 5.4$\pm$1.2 & 19.9$\pm$4.5 	& 44.1$\pm$3.2 	& 95.1$\pm$6.6 	& 105.7$\pm$7.3 	& 85.3$\pm$10.2 	& $<$32 			& 12$\pm$3 		\\
SPT2319-55 & 0.82$\pm$0.05	& 5.4$\pm$1.2 & 17.5$\pm$4.4 	& 38.1$\pm$2.9 	& 49.0$\pm$6.6 	& 44.0$\pm$6.0 	& 32.8$\pm$6.4 	& $<$8 			& $<$25 			\\
SPT2335-53 & 	0.30$\pm$0.04	& 4.9$\pm$1.3 & 13.8$\pm$3.7 	& 29.7$\pm$5.7 	& 78.6$\pm$9.9 	& 64.6$\pm$8.4 	& 61.4$\pm$9.0 	&   				&   				\\
SPT2340-59 & 	0.49$\pm$0.05	& 3.6$\pm$1.0 & 15.3$\pm$3.7 	& 34.2$\pm$4.1 	& 71.1$\pm$8.7 	& 66.1$\pm$6.9 	& 41.6$\pm$8.5 	& $<$29 			& $<$8 			\\
SPT2344-51 & $<$0.15		& 0.7$\pm$1.2 & 2.8$\pm$6.5		& 28.4$\pm$5.0	& 76.8$\pm$10.6	& 53.0$\pm$9.0	& 40.1$\pm$8.8	&				&				\\
SPT2349-50 & 	0.51$\pm$0.05	& 5.2$\pm$1.1 & 24.6$\pm$5.0 	& 42.6$\pm$3.3 	& 127.8$\pm$7.6 	& 135.8$\pm$7.1 	& 129.2$\pm$8.6 	& $<$26		& $<$13 		\\
SPT2349-56 & 	0.40$\pm$0.05	& 4.7$\pm$1.2 & 21.1$\pm$4.2 	& 56.5$\pm$8.0 	& 85.4$\pm$6.4 	& 72.4$\pm$5.9 	& 36.8$\pm$6.4 	& $<$33 			& $<$12 			\\
SPT2351-57 & 	0.83$\pm$0.05	& 5.6$\pm$1.3 & 15.7$\pm$6.3 	& 34.6$\pm$3.1 	& 73.8$\pm$5.7 	& 56.0$\pm$6.4 	& 44.3$\pm$5.3 	& $<$44 			& $<$10 			\\
SPT2353-50 & 	0.89$\pm$0.05	& 5.4$\pm$1.4 & 21.1$\pm$4.3 	& 40.6$\pm$3.8	 & 56.2$\pm$7.1 	& 51.8$\pm$6.0 	& 29.9$\pm$7.4 	& $<$41 			& $<$12			\\
SPT2354-58 & 	0.61$\pm$0.08	& 2.7$\pm$1.2 & 13.5$\pm$6.2 	& 66.0$\pm$5.1 	& 277.7$\pm$7.9 	& 469.0$\pm$9.0 	& 613.5$\pm$10.8 	& 532$\pm$59		& 239$\pm$11 		\\
SPT2357-51 & 	0.42$\pm$0.04	& 4.1$\pm$0.9 & 20.4$\pm$4.4 	& 53.4$\pm$5.4 	& 122.9$\pm$7.5 	& 112.1$\pm$6.2 	& 70.9$\pm$5.1 	& $<$34 			& $<$8 			\\
\enddata
\tablecomments{The uncertainties do not include absolute calibration errors. The 2\,mm and 1.4\,mm SPT flux densities are deboosted.}
\end{deluxetable}

\section{The Spectral Energy Distribution of a non-detection}
\label{sect:SED_non-detection}

\begin{figure*}[htb]
	\centering
	\includegraphics[viewport= 0 0 500 360, clip=true,width=8cm,angle=0]{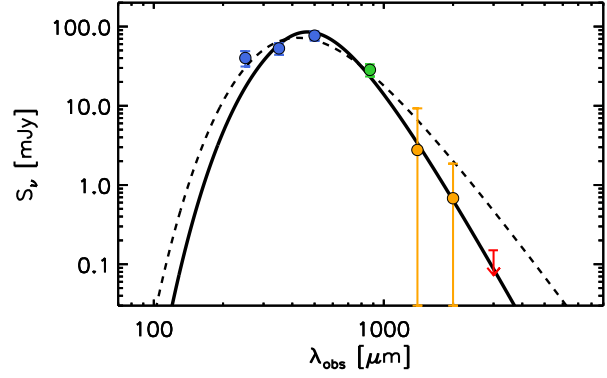}
	\caption{The Spectral Energy Distribution of SPT2344-51 with $\beta$=2 (dashed line) and $\beta$=4 (solid line). The data points are \emph{Herschel}/SPIRE at 250, 350 and 500\micron\ (\emph{blue}), APEX/LABOCA at 870\micron\ (\emph{green}), SPT at 1.4\,mm and 2.0\,mm (\emph{yellow}) and in \emph{red} is shown the 3 $\sigma$ ALMA 3\,mm detection limit.}
	\label{Fig:SED_non-detection}
\end{figure*}

Figure \ref{Fig:SED_non-detection} shows the SED of SPT2344-51. 
The photometry of this source indicated that this high-redshift source would be bright enough for detection at our 3\,mm sensitivity, but we did not detect it. 
Since the ALMA Cycle 1 deadline we detected this source with APEX/LABOCA and using this point with $\beta$=2, which is what we use in our SED fits when finding photometric redshifts (dashed line), the SED shows that we should detect the source at 3\,mm. When we instead force the SED to go through the SPT points ($\beta$=4) then the SED falls below the detection limit. This is however, a steeper slope than we would expect. This upper limit is assuming an unresolved source. 
If the source is smeared over two or more beams the flux density per beam could easily be too low to be detected.

\end{document}